\documentclass[twocolumn]{aastex63}
\usepackage{comment}
\usepackage{amsmath}
\usepackage{rotating}
\usepackage{layouts}
\usepackage{tasks}
\usepackage{enumitem}
\usepackage{placeins}
\usepackage{lineno}
\usepackage{amsmath} 
\usepackage{gensymb}

\begin{document}

\newcommand{\nopar}{{\parfillskip=0pt \par}}

\newcommand{\Mjup}{\rm{M}_{\rm{Jup}}\xspace}
\newcommand{\Rjup}{\rm{R}_{\rm{Jup}}\xspace}
\newcommand{\HD19467}{HD~19467}
\newcommand{\HD19467B}{HD~19467~B}

\newcommand{\todo}[1]{\textcolor{red}{[#1]}}

\shorttitle{Spectroscopy of HD 19467 B with the NIRSpec IFU}
\shortauthors{Hoch et al.}

\title{JWST-TST High Contrast: Spectroscopic Characterization of the Benchmark Brown Dwarf HD 19467 B with the NIRSpec Integral Field Spectrograph}

\author[0000-0002-9803-8255]{Kielan K. W. Hoch}
\affiliation{Space Telescope Science Institute, 3700 San Martin Dr, Baltimore, MD 21218, USA}

\author[0000-0002-9807-5435]{Christopher A. Theissen}
\affiliation{Department of Astronomy \& Astrophysics, University of California, San Diego, La Jolla, CA 92093, USA}

\author[0000-0002-7129-3002]{Travis S. Barman}
\affiliation{Lunar and Planetary Laboratory, University of Arizona, Tucson, AZ 85721, USA}

\author[0000-0002-3191-8151]{Marshall D. Perrin}
\affiliation{Space Telescope Science Institute, 3700 San Martin Dr, Baltimore, MD 21218, USA}

\author[0000-0003-2233-4821]{Jean-Baptiste Ruffio}
\affiliation{Department of Astronomy \& Astrophysics,  University of California, San Diego, La Jolla, CA 92093, USA}


\author[0000-0003-4203-9715]{Emily Rickman}
\affiliation{European Space Agency (ESA), ESA Office, Space Telescope Science Institute, 3700 San Martin Dr, Baltimore, MD 21218, USA}

\author[0000-0002-9936-6285]{Quinn M. Konopacky}
\affiliation{Department of Astronomy \& Astrophysics,  University of California, San Diego, La Jolla, CA 92093, USA}

\author[0000-0003-0192-6887]{Elena Manjavacas}
\affiliation{AURA for the European Space Agency (ESA), ESA Office, Space Telescope Science Institute, 3700 San Martin Drive, Baltimore, MD, 21218 USA}
\affiliation{Department of Physics and Astronomy, Johns Hopkins University, Baltimore, MD 21218, USA}

\author[0000-0001-6396-8439]{William O. Balmer}
\affiliation{ Department of Physics \& Astronomy, Johns Hopkins University, 3400 N. Charles Street, Baltimore, MD 21218, USA}
\affiliation{Space Telescope Science Institute, 3700 San Martin Dr, Baltimore, MD 21218, USA}

\author[0000-0003-3818-408X]{Laurent Pueyo}
\affiliation{Space Telescope Science Institute, 3700 San Martin Dr, Baltimore, MD 21218, USA}

\author[0000-0003-2769-0438]{Jens Kammerer}
\affiliation{European Southern Observatory, Karl-Schwarzschild-Str. 2, 85748 Garching, Germany}
\affiliation{Space Telescope Science Institute, 3700 San Martin Dr, Baltimore, MD 21218, USA}

\author[0000-0001-7827-7825]{Roeland P. van der Marel}
\affiliation{Space Telescope Science Institute, 3700 San Martin Dr, Baltimore, MD 21218, USA}
\affiliation{Center for Astrophysical Sciences, The William H. Miller III Department of Physics \& Astronomy, Johns Hopkins University, Baltimore, MD 21218, USA}

\author[0000-0002-8507-1304]{Nikole K. Lewis} 
\affiliation{Department of Astronomy and Carl Sagan Institute, Cornell University, 122 Sciences Drive, Ithaca, NY 14853, USA}

\author[0000-0001-8627-0404]{Julien H. Girard}
\affiliation{Space Telescope Science Institute, 3700 San Martin Dr, Baltimore, MD 21218, USA}

\author[0000-0002-6892-6948]{Sara~Seager}
\affiliation{Department of Physics and Kavli Institute for Astrophysics and Space Research,  Massachusetts Institute of Technology, Cambridge, MA 02139, USA}
\affiliation{Department of Earth, Atmospheric and Planetary Sciences, Massachusetts Institute of Technology, Cambridge, MA 02139, USA}
\affiliation{Department of Aeronautics and Astronautics, MIT, 77 Massachusetts Avenue, Cambridge, MA 02139, USA}

\author[0000-0003-4003-8348]{Mark Clampin}
\affiliation{Astrophysics Division, Science Mission Directorate, NASA Headquarters, 300 E Street SW, Washington, DC 20546, USA}

\author{C. Matt Mountain}\affil{Association of Universities for Research in Astronomy, 1331 Pennsylvania Avenue NW Suite 1475, Washington, DC 20004, USA}  

\correspondingauthor{Kielan K. W. Hoch}
\email{khoch@stsci.edu}


\begin{abstract}
We present the atmospheric characterization of the substellar companion HD 19467 B as part of the pioneering JWST GTO program to obtain moderate resolution spectra (R$\sim$2,700, 3-5$\mu$m) of a high-contrast companion with the NIRSpec IFU. HD 19467 B is an old, $\sim$9 Gyr, companion to a Solar-type star with multiple measured dynamical masses. The spectra show detections of CO, CO$_2$, CH$_4$, and H$_2$O. We forward model the spectra using Markov Chain Monte Carlo methods and atmospheric model grids to constrain the effective temperature and surface gravity. We then use NEWERA-PHOENIX grids to constrain non-equilibrium chemistry parameterized by $K_{zz}$ and explore molecular abundance ratios of the detected molecules. We find an effective temperature of 1103 K, with a probable range from 1000--1200 K, a surface gravity of 4.50 dex, with a range of 4.14--5.00, and deep vertical mixing, log$_{10}$($K_{zz}$), of 5.03, with a range of 5.00--5.44. All molecular mixing ratios are approximately Solar, leading to a C/O $\sim$0.55, which is expected from a T5.5 brown dwarf. Finally, we calculate an updated dynamical mass of HD 19467 B using newly derived NIRCam astrometry which we find to be $71.6^{+5.3}_{-4.6} M_{\rm{Jup}}$, in agreement with the mass range we derive from evolutionary models, which we find to be 63-75 $M_{\rm{Jup}}$.These observations demonstrate the excellent capabilities of the NIRSpec IFU to achieve detailed spectral characterization of substellar companions at high-contrast close to bright host stars, in this case at a separation of $\sim$1.6\arcsec with a contrast of 10$^{-4}$ in the 3-5 $\mu$m range.
\end{abstract}

\section{Introduction}\label{intro}
The James Webb Space Telescope (JWST) has opened the door to spectroscopy of directly imaged exoplanets and brown dwarfs beyond 3 \micron. This wavelength regime probes critical molecular species from the atmospheres of cool, substellar objects, yielding new insights into the atmospheric processes of these Jupiter-like companions.  Species such as H$_2$O, CO, CO$_2$, and CH$_4$ have been detected in transiting planets, free-floating brown dwarfs, and directly imaged companions \citep{cushing2005,birkby2013,gonzales2021,konopacky2013}, meaning that such detections are feasible for companions given sufficient signal-to-noise at moderate spectral resolution for further characterization. 


\par Achieving high fidelity molecular detections in this wavelength regime (3-5$\mu$m) will require applying JWST's exquisite spectroscopic sensitivity to substellar companions seen in the glare of their host stars, including both planetary and brown dwarf companions. Brown dwarfs have similar spectral features to directly imaged planets, and can be used as test cases for verifying modeling procedures that characterize substellar atmospheres. Many brown dwarfs also have existing constraints on effective temperature, surface gravity, and composition. Brown dwarfs that are companions to main-sequence stars are especially useful because they may inherit stellar properties such as metallicity and share similar ages to their host star. For our program, we aim to show that we can obtain the composition that we expect for a massive brown dwarf companion to be confident in our future analyses of exoplanets. Furthermore, cases in which the companion's mass can be measured dynamically from mapping their orbits---often referred to as ``benchmark" systems---are especially valuable for testing evolutionary models for changes in interior properties and atmospheric properties over cosmic time. The growing list of benchmark brown dwarfs include HIP 64892 \citep{cheetham2018}, GL 758 B \citep{bowler2018}, HD 19467 B \citep{maire2020}, and more recently HD 112863 B and HD 206505 B \citep{rickman2024}. These objects are essential to understand low-mass object evolution.

This paper is part of a series by the JWST Telescope Scientist Team (JWST-TST; PI Matt Mountain)\footnote{\url{https://www.stsci.edu/~marel/jwsttelsciteam.html}}. In addition to providing scientific support for observatory development through launch and commissioning, the team was awarded 210 hours of Guaranteed Time Observer (GTO) time, which is being used for studies in three different subject areas: (a) Transiting Exoplanet Spectroscopy (lead: N. Lewis, e.g. \citealt{grant2023}); (b) Exoplanet and Debris Disk High-Contrast Imaging (lead: M. Perrin, e.g. \citealt{rebollido2024}); and (c) Local Group Proper Motion Science (lead: R. van der Marel, e.g. \citealt{libralato2023}). A common theme of these investigations is the desire to pursue and demonstrate science for the astronomical community at the limits of what is made possible by the exquisite optics and stability of JWST.  

The program presented here was designed to test and demonstrate methods for high contrast imaging spectroscopy with the NIRSpec IFU, and for wider application of these techniques to larger samples in subsequent cycles. 
These JWST observations of HD 19467 B were the first high contrast imaging spectroscopy observations using the NIRSpec Integral Field Unit (IFU) \citep{jakobsen2022,boker2022}, and demonstrate the capabilities of the instrument for in-depth spectral characterization of low-mass high contrast substellar companions. HD 19467 B has an apparent angular separation of 1.6 arcseconds, and a contrast of $\sim10^{-4}$ in the Near-Infrared (NIR), making it a feasible but challenging target for spectroscopy using the NIRSpec IFU. 

Moderate resolution integral field spectroscopy has made it possible to disentangle host starlight from their cooler, fainter companions to resolve distinct molecular features to study their atmospheres. This has been done from the ground with instruments such as OSIRIS on the W.M Keck Telescope \citep{larkin2006}, and SINFONI on the VLT \citep{eisenhauer2003,bonnet2004}. IFU observations have revealed molecular features from species such as CO and H$_2$O in the $J$-, $H$-, $K$-band from the ground \citep{konopacky2013,barman2015,petitditdelaroche2018, ruffio2019, wilcomb2020,ruffio2021,petrus2021,hoch2022,hoch2023}. These molecular features have been used to derive carbon-to-oxygen (C/O) ratios that have been postulated to trace how and where the companions formed relative to their host stars \citep[e.g.][]{oberg2011,molliere2022,hoch2023}. With the NIRSpec IFU, \cite{miles2023} presented the first NIRSpec IFU spectra of a low contrast substellar companion, VHS 1256 b \citep{gauza2015}, revealing the molecular detection of CH$_4$ in addition to H$_2$O and CO from the moderate resolution NIRSpec observations alone. Our JWST NIRSpec data of HD 19467 B has put ground-based high contrast reduction and analysis methods to the test in order to extract a spectrum of this faint companion.

In a companion paper, \citet{Ruffio2023}, we present the data reduction and analysis methods for high contrast imaging spectroscopy using the NIRSpec IFU as developed and applied to this dataset. In particular we use forward modeling of the measured detector images as a key strategy to minimize systematics, along with point spread function (PSF) subtraction using reference differential imaging to yield a high-quality, high-contrast spectrum of HD 19647 B.

In this paper, we characterize the atmosphere of HD 19467 B using that spectrum, including an in-depth analysis of the spectral features, molecular abundances and ratios, and show the level of constraints that this new wavelength regime allows. HD 19467 B is among the subset of substellar companions for which orbital accelerations allow inference of the dynamical mass.  We present an updated dynamical mass and new estimates of the predicted mass from evolutionary models, or model-dependent mass, to test whether the masses are consistent.  




In the remainder of this Introduction we summarize key properties of the HD 19467 system known from prior work.  In Section \ref{obs}, we summarize the observations taken with the NIRSpec IFU. In Section \ref{data_reduction}, we summarize the data reduction and analysis used to extract the spectra as already described in depth in \citet{Ruffio2023}. We briefly explore the broad absorption spectral features that can be seen in the data in Section \ref{spec_features}. To validate these features, we conduct an empirical analysis in comparison with AKARI brown dwarf spectra in Section \ref{emp_analysis}. Section \ref{forward_modeling} steps through how we forward model the data using three large atmospheric model grids and a new PHOENIX model grid to explore non-equilibrium chemistry. In Section \ref{molecules} we create smaller PHOENIX grids that vary molecular abundances to constrain abundance ratios. We also measure an updated dynamical mass and compare it to evolutionary models in Section \ref{dyn_mass}. Next, we discuss our findings and results in Section \ref{discussion} and conclude our work in Section \ref{conclusion}.

\subsection{Prior studies of the HD 19467 system}\label{hd19467b}

\begin{figure*}
\centering
\includegraphics[width=0.49\textwidth]{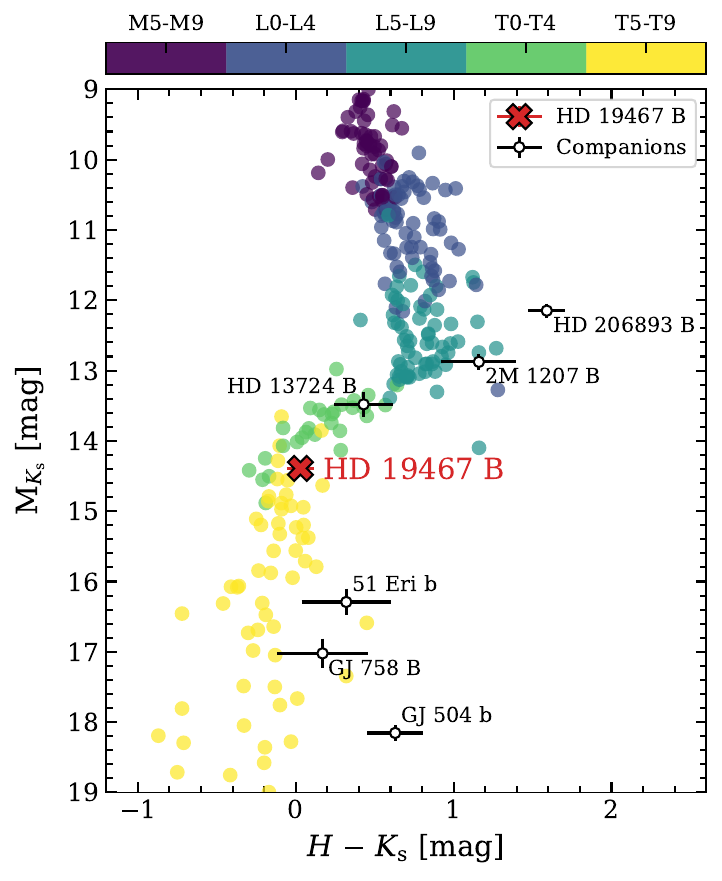}
\includegraphics[width=0.49\textwidth]{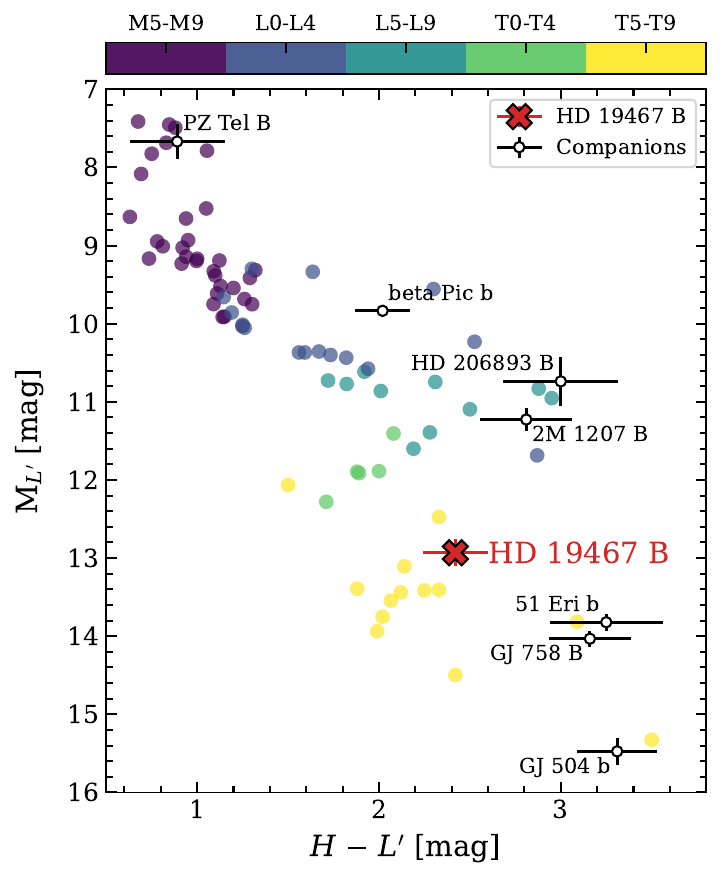}
\caption{Color-magnitude diagrams illustrating where HD 19467 B lies compared to various directly imaged companions (shown in white circles), and field brown dwarfs (shown in filled-in circles). The field brown
dwarfs are color-coded by spectral classification. Photometry from directly imaged companions taken from: \cite{kammerer2021}, \cite{chauvin2004}, \cite{rickman2020}, \cite{maire2020}, \cite{maire2019}, \cite{vigan2016}, \cite{janson2013}, \cite{bonnefoy2011}, \cite{biller2010} and plotted using \texttt{species} \cite{stolker2020}.}
\label{fig:cmd}
\end{figure*}


\par HD 19467 is a G3V star at a distance of 32.03$\pm$0.03 pc \citep{gaia2021}, with a T5 companion at $\sim$1.6'' discovered via the radial velocity (RV) technique and confirmed via coronagraphic imaging with Keck NIRC2 \citep{crepp2014}. 

There are several lines of evidence that this is a relatively old star, though the range of estimates is large. \cite{wood2019} quoted the age estimate of HD 19467 to be 10.06$^{+1.16}_{-0.82}$ Gyr with isochronal models. \cite{maire2020} estimated an age of 8.0$^{+2.0}_{-1.0}$ Gyr via isochrones, gyrochronology, and chemical and
kinematic arguments. \cite{gomesdasilva2021} found an age of 11.882$\pm2.564$ Gyr via isochrone fitting, but \cite{brandt2021} places an activity-age at 5.41$^{+1.8}_{-1.34}$ Gyr. The most recent age estimate is from \cite{greenbaum2023} and they place the age at 9.4$\pm1.0$ Gyr using asteroseismic detection from TESS, which we adopt for this work. A summary of the system parameters are shown in Table \ref{tab:sum_system}. 


\par As stated above, HD 19467 B was first imaged by \cite{crepp2014} using Keck NIRC2 to obtain photometry in $J$-, $H$-, and $K_{s}$-band. \cite{crepp2014} found that the colors and absolute magnitudes were consistent with a late T-dwarf (T5-T7). In a follow up paper, \cite{crepp2015} obtained P1640 low resolution (R$\sim$30) spectra of the companion across $J$- and $H$- bands and fit BTSETTL models to find an effective temperature of 978$^{+20}_{-43}$ K and a spectral type of T5.5$\pm$1. \cite{maire2020} revisited HD 19467 B with SPHERE and NaCo on the VLT to obtain more photometry across $H2$-,$H3$-.$K1$-,$K2$-,$L'$-, and $M'$-bands. \cite{maire2020} found an effective temperature of 1042$^{+77}_{-71}$ K by modeling the photometric data available on the object. \cite{mesa2020} obtained IRDIS long-slit spectroscopy at a resolution of R$\sim$ 350 of HD 19467 B and fit BTSETTL models to obtain an effective temperature of 1000$\pm$100 K and a surface gravity of 5$\pm$0.5 dex, with a spectral type of T6$\pm$1. 



\par A key aspect of this system, and a significant motivating factor for its choice as a GTO target, is the availability of a dynamical mass measurement for the companion. Most directly imaged substellar companions only have evolutionary model-dependent mass estimates. The first estimate of the dynamical mass of the brown dwarf HD~19467~B initially had a lower limit derived from the host star radial velocities of M$\geq51.9^{+3.6}_{-4.3}M_{\mathrm{Jup}}$ \citep{crepp2014,crepp2015}. Additional relative astrometry from high-contrast imaging with VLT/SPHERE \citep{maire2020} were added into the orbital fit of HD~19467~B, giving a dynamical mass of 74$^{+12}_{-9}~M_{\mathrm{Jup}}$ \citep{maire2020}, right below the hydrogen-burning limit. Subsequently, \cite{brandt2021} revisited this orbital fit and incorporated Hipparcos and \emph{Gaia} eDR3 astrometry giving a dynamical mass of $65.4^{+5.9}_{-4.6}~M_{\mathrm{Jup}}$. Most recently, \cite{greenbaum2023} estimated a dynamical mass of 81$^{+14}_{-12}~M_{\mathrm{Jup}}$ using new relative astrometry from JWST's NIRCam as well as radial velocities from HARPS and HIRES. This estimate is high compared to the model-derived mass of 62$\pm$1$M_{\mathrm{Jup}}$ from \cite{greenbaum2023}, which is closer to values from \cite{brandt2021} and \cite{maire2020}, but still lower.
This tension between the dynamical and evolutionary model-based masses motivates further investigation of this system. In this paper, we conduct our analysis using updated astrometry from \citet{Ruffio2023} and find a dynamical mass of 71.6$^{+5.3}_{-4.6}M_{\mathrm{Jup}}$, which we discuss in Section \ref{dyn_mass}. Furthermore, we carry out a comparison between the calculated dynamical mass and the model-derived mass from evolutionary models which is discussed in Section \ref{evol_model}.

Figure \ref{fig:cmd} shows the relative position of HD 19647B on color-magnitude diagrams, demonstrating its moderately bright magnitude at relevant wavelengths, but also its overlap in color space with other directly imaged companions.

\par In addition to HD 19467 B having dynamical mass constraints, it is a favorable target due to its contrast and separation relative to the host star. \cite{crepp2014} found a contrast ratio of 9.4$\times10^{-6}$ in the $K_s$-band from the ground. From space with JWST, \cite{greenbaum2023} found the companion detectable with contrasts 2$\times10^{-4}$--2$\times10^{-5}$ using all 6 filters of NIRCam with ADI and Synthetic Reference Differential Imaging (SynRDI).   HD 19467 B therefore represents an ideal case to test High Contrast Imaging with the NIRSpec IFU given its moderate contrast in the NIR. 



\begin{deluxetable}{lccc} 
\tabletypesize{\scriptsize} 
\tablewidth{0pt}
\tablecaption{Summary of HD 19467 B System \label{tab:sum_system}} 
\tablehead{ 
 \colhead{Parameter} & \colhead{Host Star} & \colhead{Companion} & \colhead{\textbf{Reference}}}
\startdata 
Distance (pc) & 32.03$\pm 0.03$ &  - & \tablenotemark{a} \\
Separation (arcsec) & - &  $1.640 \pm 0.007$ & \tablenotemark{d} \\
Spectral Type & G3V & T5.5$ \pm 1.0$ & \tablenotemark{c,d}\\
Age (Gyr) & 9.4$\pm1.0$ & - & \tablenotemark{e} \\
Mass (M$_\odot$, M$_\mathrm{Jup}$) & 0.96$\pm0.2$ & $71.6^{+5.3}_{-4.6}$ & \tablenotemark{e}, \tablenotemark{f}
\enddata
\tablerefs{(a) \cite{gaia2021}, (b)  \cite{crepp2014}, (c)  \cite{gomesdasilva2021}, (d) \cite{crepp2015}, (e) \cite{greenbaum2023}, (f) this work}
\end{deluxetable}

\section{Observations}\label{obs}

We  summarize succinctly the observational details already presented in \citet{Ruffio2023}. 

\subsection{NIRSpec IFU Observations}
HD 19467 was observed with the JWST NIRSpec IFU on 2023 January 25 (UT) in observing program 1414.
The F290LP filter was chosen for the wavelength range 2.9--5.3 $\mu$m, with the grating G395H for the highest available resolution (R$\sim$2700). HD 18511 was chosen to be the PSF reference star for these observations. Both HD 19467 and HD 18511 were too bright for target acquisition, so absolute pointing was used with an accuracy of $0.1''$. A 9-point small cycling dither pattern was chosen to help with the spatial undersampling of NIRSpec while keeping the companion in the field of view (FOV). The NRSIRS2RAPID readout pattern was used with 15 groups per integration for HD 19467 (yielding a total exposure time on source of $\sim35$ minutes), and 5 groups for the brighter HD 18511.  

The originally-intended observing plan included a suite of observations for testing and comparing several differential imaging strategies for high contrast; however due to guide star acquisition issues only a portion of that plan was successfully obtained. The successful observations, summarized in Table \ref{tab:obs}, include one science observation and its corresponding PSF reference observation (plus an additional PSF observation with an offset pointing, for which there is no corresponding science image). This partial dataset suffices for Reference Differential Imaging (RDI), and so we restrict our approach to only that method.



\begin{deluxetable*}{lccrccl} 
\tabletypesize{\scriptsize} 
\tablewidth{0pt} 
\tablecaption{Summary of NIRSpec IFU F290LP/G395H Observations.}
\label{tab:obs}
\tablehead{ 
  \colhead{Obs.} & \colhead{Target} &  \colhead{Groups} & $T_{\mathrm{int}}$ per exp. & Dithers & Total $T_{\mathrm{int}}$&
  \colhead{Description}} 
\startdata
1 & HD 18511 & 5 &  87.5 s & 9 & 13 min. & Reference star off FOV (not used) \\
4 & HD 19467 & 15 & 233.4 s & 9 & 35 min. & Science star in center of FOV at roll angle 2 \\
5 & HD 18511 & 5 &  87.5 s & 9 & 13 min. & Reference star in center of FOV\\
\enddata
 \tablenotetext{~}{All exposures use NRSIRS2RAPID readout pattern, 1 integration per dither. This data can be found in MAST: \dataset[https://doi.org/10.17909/q524-zn59]{https://doi.org/10.17909/q524-zn59}.}
\end{deluxetable*}

\subsection{NIRCam Observations}

As previously reported by \citet{greenbaum2023}, HD~19467 and its brown dwarf companion were observed with JWST/NIRCam \citep{rieke2003,rieke2023} bar mask coronagraphy on 2022 August 12 (program 1189; PI Thomas Roellig) as part of the NIRCam GTO program. These observations included six filters in NIRCam's 
long-wavelength channel with nominal wavelengths from 2.5 to 4.6 \micron, largely overlapping with the NIRSpec F290LP bandpass. Unfortunately, due to a target acquisition issue with the reference star, no PSF reference observations were obtained. This limits the analysis to angular differential imaging methods using the two observed roll angles, or alternative methods to simulate or empirically estimate a PSF reference, both of which were explored in \citet{greenbaum2023}.

We described in \citet{Ruffio2023} that we performed a re-reduction of those NIRCam data using updated pipelines and calibration files, in particular employing the \texttt{spaceKLIP}\footnote{\url{https://github.com/kammerje/spaceKLIP}} community pipeline \citep{Kammerer2022,Carter2023} for JWST coronagraphy data and its implementation of angular differential imaging \citep{marois2006} using the KLIP algorithm \citep{soummer2012}. This re-reduction also benefited from more recent in-flight flux calibrations for NIRCam coronagraphy, which were not yet available during the initial reduction by \citet{greenbaum2023}. From that re-reduction we extracted updated photometry and uncertainties for the companion HD~19467~B from 2.5 to 4.6~\micron~using PSF forward modeling within \texttt{pyKLIP} \citep{wang2015,pueyo2016}. Our updated NIRCam photometry of the brown dwarf companion including estimated uncertainties can be found in Table~\ref{tab:photometry}.

\begin{table*}[!ht]
    \begin{tabular}{l c c c c c c}
        Filter & Flux & Flux& Systematic error& $\Delta\rm{mag}$ & $\rm{TP}_{\rm{MSK}}$ & $\rm{TP}_{\rm{COM}}$ \\
         & ($\mu$Jy) & ($10^{-17}\,\mathrm{W}/\mathrm{m}^2/\mu\mathrm{m}$)& ($10^{-17}\,\mathrm{W}/\mathrm{m}^2/\mu\mathrm{m}$) & & & \\
        \hline
        F250M & $15.6\pm1.1$ & $0.74\pm0.05$ &  0.04 (5\%) & $13.41\pm0.08$ & 1.000 & 0.963\\
        F300M & $31.5\pm0.3$ & $1.06\pm0.01$ &  0.05 (5\%) & $12.33\pm0.01$ & 1.000 & 0.898\\
        F360M & $62.4\pm0.5$ & $1.42\pm0.01$ &  0.07 (5\%) & $11.21\pm0.01$ & 0.999 & 0.951\\
        F410M & $187.1\pm1.3$ & $3.37\pm0.02$ &  0.17 (5\%) & $9.78\pm0.01$ & 0.966 & 0.958\\
        F430M & $125.7\pm1.1$ & $2.06\pm0.02$ &  0.10 (5\%) & $10.10\pm0.01$ & 0.943 & 0.947\\
        F460M & $84.4\pm1.0$ & $1.18\pm0.01$ &  0.06 (5\%) & $10.33\pm0.01$ & 0.918 & 0.901\\
        \hline
    \end{tabular}
    \caption{Revised \emph{JWST}/NIRCam photometry of HD~19467~B with statistical uncertainties from the MCMC fit of the forward-modeled NIRCam PSF to the data. The systematic uncertainties are estimated to be $\sim10\%$ \citep[see][]{Ruffio2023}. $\rm{TP}_{\rm{MSK}}$ and $\rm{TP}_{\rm{COM}}$ denote the throughput of the coronagraphic mask and the \textbf{Coronagraphic Occulting Mask} (COM) substrate, respectively.}
    \label{tab:photometry}
\tablenotetext{*}{These filters partially overlap with the wavelength gap between the two detectors in NIRSpec. The photometry was therefore computed after interpolating the spectrum in the gap with a BT-Settl model.}
\end{table*}

\section{Data Reduction and Spectral Extraction}\label{data_reduction}
The spectral extraction of a high-contrast companion like HD~19467~B with the NIRSpec IFU is challenging and requires dedicated tools to manage NIRSpec's spatial undersampling. We found an effective approach is to perform the stellar PSF subtraction and the flux extraction directly in detector images instead of spectral cubes. This was the topic of \citet{Ruffio2023}, which analyzed the same dataset; we use the spectrum extracted therein. We summarize the effects of reconstructed spectral cubes and our subsequent analysis below, and encourage readers to consult that paper for a more detailed presentation of the method.

\citet{Law2023} illustrated the negative effects of the spatial undersampling on the reconstructed spectral cubes in the context of MIRI MRS. Indeed, spectra of individual spaxels feature large oscillations up to 50\% of their continuum. The effect can be reduced by increasing the number of dithers, e.g. halved with four dithers. These effects are similar for the NIRSpec IFU preventing an adequate spectral extraction for HD~19467~B from the spectral cubes.

The reduction was started using uncalibrated NIRSpec files. We used the science calibration pipeline version "1.10.2.dev7+g8fb5bd7d"\footnote{This is a Git version tag, representing running the latest-available development pipeline at the time of code execution.} stages 1-2 to generate flux calibrated detector images \citep{Bushouse2023}. Once we obtained the calibrated files, we used \texttt{breads}\footnote{\url{https://github.com/jruffio/breads}} to identify and mask bad pixels not flagged in the calibration pipeline, interpolate the wavelengths onto an evenly sampled wavelength grid, recalculate the WCS RA and Dec. coordinates to arcseconds, and to mask the area of the detector affected by charge diffusion between pixels due to the saturation of the star on the detectors.


To obtain the spectrum of the companion, \citet{Ruffio2023} subtracted the stellar PSF using an implementation of reference-star differential imaging (RDI) that can be directly applied to NIRSpec detector images after they were interpolated on a regularly sampled wavelength grid, but without spatial interpolation. The dedicated observation of the reference star HD~18511 was used for RDI (Obs. 5 in Table \ref{tab:obs}). As mentioned previously, the detector images are interpolated in the wavelength direction to produce monochromatic slices of the 3D data. The resulting collection of pixels at a constant wavelength is a irregular sampling of the field of view, which is referred to as the point cloud. PSF fitting is then performed in this point cloud at each wavelength. The stellar PSF is subtracted by interpolating the reference star point cloud onto the science point cloud and fitting for a scaling factor and centroid offset.

\begin{figure*}
\centering
  \includegraphics[width=\textwidth]{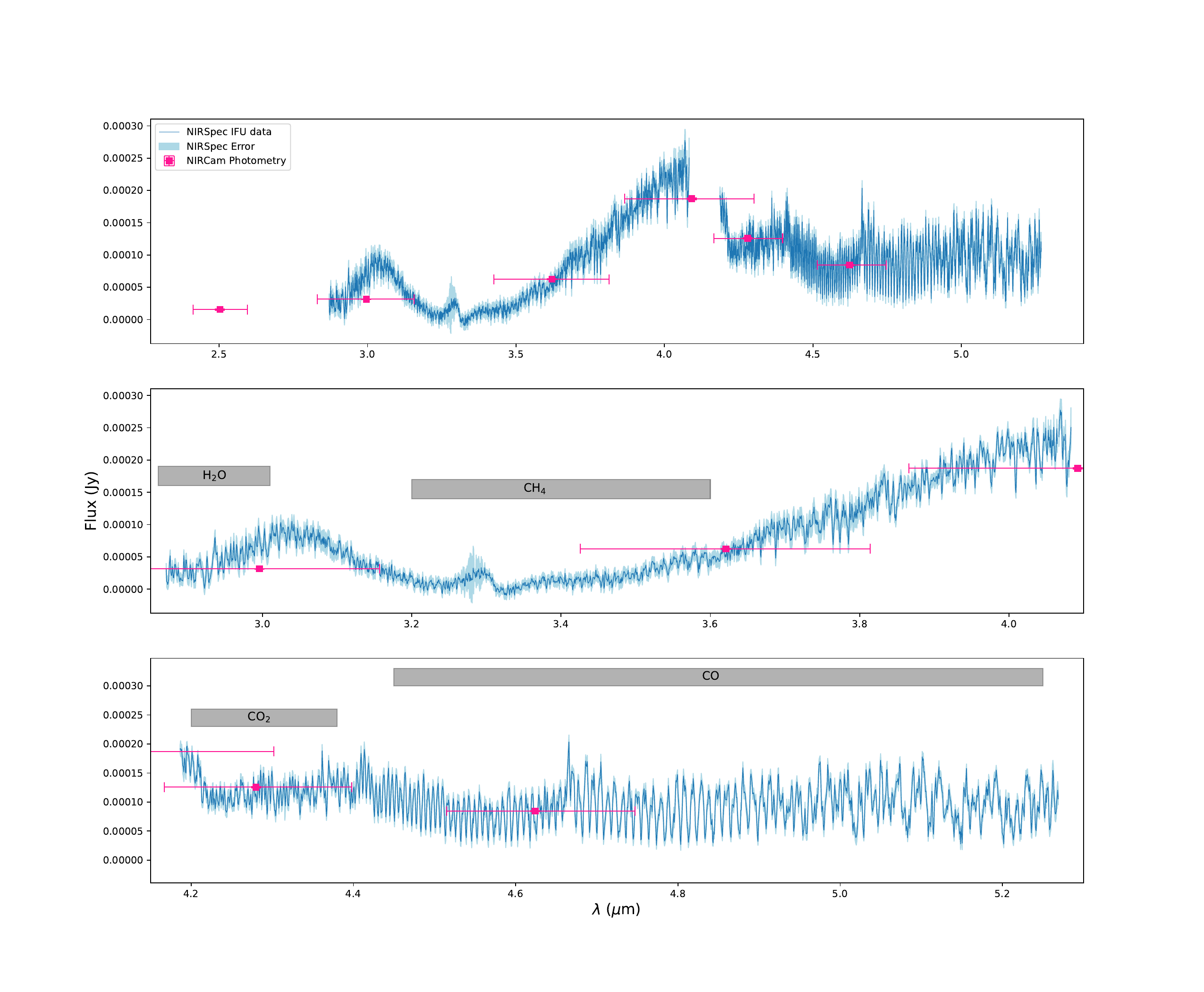}
\caption{JWST NIRSpec IFU G395H/F290LP R$\sim$2,900 spectrum of HD 19467 B spanning 2.9--5.3 $\mu$m, obtained for a companion at a contrast of $\sim10^{-4}$ and separation of 1.6 arcsec relative to the host star. The custom data analysis developed to process and extract this spectrum are summarized in Section \ref{obs} and a more detailed procedure can be found in \citet{Ruffio2023}. A cursory molecular inventory in this NIRSpec spectrum indicates the presence of H$_2$O, CH$_4$, CO and CO$_2$, which are labeled in gray above their corresponding wavelength ranges. The spectrum is in dark blue, the error is in light blue, and NIRCam photometry from our re-reduction of the \citet{greenbaum2023} observations are shown in pink. The top panel shows the entirety of JWST data for HD 19467 B. The middle panel focuses in on the first detector wavelength range of NIRSpec (2.9--4.1 $\mu$m), and the third panel shows the second detector wavelength range of NIRSpec (4.2--5.3 $\mu$m).}
\label{fig:spec}
\end{figure*}

After the PSF subtraction, the spectrum of HD~19467~B was extracted by fitting a WebbPSF model to the speckle-subtracted point cloud at each wavelength. This method significantly reduces the effects of the spatial undersampling and resulted in the spectra shown in figure \ref{fig:spec}. However, \citet{Ruffio2023} noted that the companion spectrum remained limited by the residual speckles due to interpolation errors of the reference star onto the science sampling. The speckle interpolation error corresponds to a first order approximation of at least 50\% of the total RDI error. We point the reader to Section E of the Appendix in \cite{Ruffio2023} for an in depth discussion about correlated uncertainties in the spectrum from interpolation noise on small spectral scales ($<2$ pixels) and the noise correlated on larger scales from residual speckles. There is ongoing work to mitigate these effects because the residuals are not a fundamental limitation of the data.


\section{Spectrum Overview}\label{overview}
\label{spec_features}
The obtained spectrum is rich in detail, with many molecular lines detected at NIRSpec's $R\sim 2700$ resolution.
Before examining the spectrum of HD 19467 B in-depth, we provide a general overview describing the features observed in the spectrum. 
The extracted NIRSpec spectra is shown in Figure \ref{fig:spec} with broadband molecular features pointed out as gray regions. 
We see evidence for absorption from H$_2$O from about 2.9--3 $\mu$m, CH$_4$ molecular features appear around 3.3~$\mu$m, CO$_2$ absorption appears around 4.2~$\mu$m, and CO absorption can be seen from 4.5 $\mu$m. These species are expected from low-temperature objects such as HD 19467 B. 

To ensure the fidelity of the absolute flux calibration we compare our extracted spectrum to the updated NIRCam photometry from \citet{Ruffio2023}. The NIRCam values are within the error bars of the NIRSpec spectra, with some points within 5\% of the error. 

\section{Empirical Analysis}\label{emp_analysis}

\begin{figure*}
\centering
  \includegraphics[width=7in]{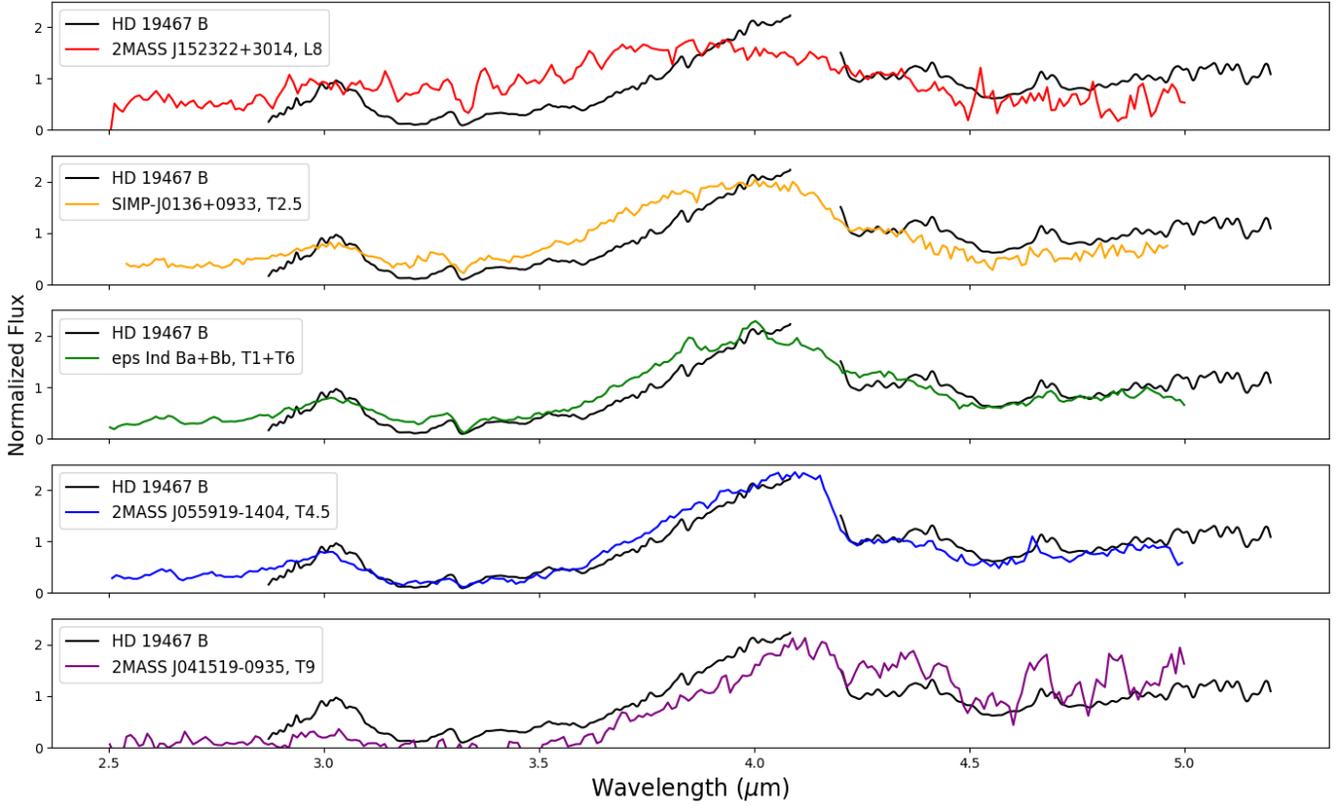}
\caption{Spectra of brown dwarfs obtained with AKARI/IRC from \cite{yamamura2010} and \cite{sorahana2013} compared to the NIRSpec IFU spectrum of HD 19467 B. The NIRSpec spectrum was down-sampled to match the resolution of the AKARI spectra ($R \approx 120$). Each panel displays the NIRSpec spectrum in black with an AKARI spectrum plotted as a colored line. The best fit AKARI spectrum is 2MASS J055919-1404, which has the lowest chi-squared value.}
\label{fig:akari}
\end{figure*}


To analyze the broad features in HD 19467 B's NIRSpec spectra, we compare our NIRSpec spectra to various brown dwarf spectra over the same wavelength range and that span L and T spectral types, using spectra obtained with the AKARI space telescope \citep{murakami2007} from \cite{yamamura2010} and \cite{sorahana2012}. 
\par The Infrared Camera (IRC; \citealt{onaka2007}) onboard AKARI conducted imaging and spectroscopic observations following its launch in 2006. The program, NIRLT (PI: I. Yamamura), obtained a set of legacy spectra of L and T dwarfs from 2.5--5.0 $\mu$m with spectral resolution $\lambda/\Delta\lambda \approx 120$. The AKARI spectra for these objects also show evidence of H$_2$O, CO$_2$, CH$_4$, and CO, similar to HD 19467 B. 
There are relatively few spectra of similar-type objects to compare to at these wavelengths as ground-based observations are limited by telluric absorption and only AKARI offers comparable spectral wavelength coverage.

\begin{figure*}[ht!]
\centering
  \includegraphics[width=7in]{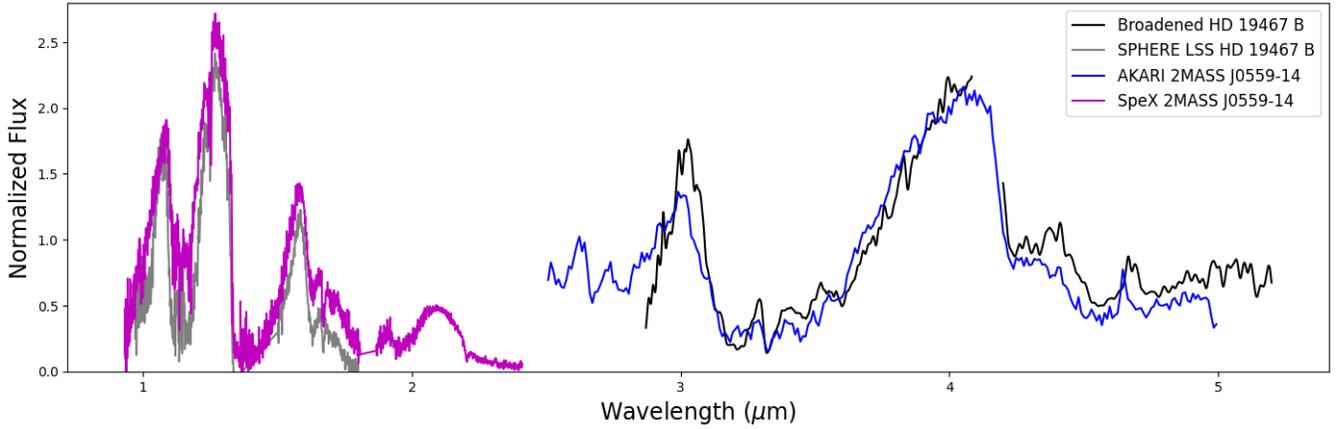}
\caption{Spectral comparison of 2MASS J0559-14 to HD 19467 B. This object is the best fit from our spectral template fitting to HD 19467 B. The broadened NIRSpec IFU data (black) and AKARI data (in blue) were converted into F$\lambda$ and then normalized. The flux-normalized SPHERE IRDIS spectra of HD 19467 B is plotted in gray and the SpeX spectra of 2MASS J0559-14 is plotted in magenta. }
\label{fig:2mass}
\end{figure*}

\par To conduct our analysis we smoothed the spectral resolution of the JWST NIRSpec spectrum to $R \approx 120$ by assuming a gaussian line spread function to match the AKARI spectral resolution, with all spectra presented in $F_\lambda$ units of W m$^{-2}$ $\mu$m$^{-1}$. We then normalized both the JWST and AKARI spectra by dividing the spectra by their mean flux value. We conducted a chi-squared analysis for each spectrum as shown in Figure~\ref{fig:akari}, and found 2MASS J055919$-$1404, a T4.5 dwarf (fourth panel down in Figure \ref{fig:akari}) to have the best-fit (lowest chi-squared value). 

\par 2MASS J055919-1404 (hereafter 2MASS J0559-14) was discovered in 2000 by \cite{burgasser2000} and was the first brown dwarf identified by the CorMASS instrument on the Palomar telescope. \cite{burgasser2006a} classified this object as a T4.5 dwarf. Its near-infrared spectra show CH$_4$ bands \citep{burgasser2000}, and \cite{yamamura2010} found evidence for deep CO and CO$_2$ absorption beyond 4.0 $\mu$m and CH$_4$ and H$_2$O bands below 3.8 $\mu$m. 

\par 2MASS J0559-14 also has a spectrum obtained with the SpeX instrument \citep{rayner2009} on the NASA Infrared Telescope Facility, that we compare here to the SPHERE spectrum from \cite{mesa2020} in Figure~\ref{fig:2mass} along with the broadened NIRSpec spectrum and AKARI spectrum.  The near-infrared data is reasonably well-matched  by the 2MASS J0559-14 data, validating its mid-infrared similarity to HD 19467B.

\par We note that \cite{crepp2015}  did a similar empirical analysis using near-IR spectra, and found the T5.5 dwarf 2MASS J11101001+0116130 \citep{burgasser2006a} was the best fit, but the lack of any 3-5 micron spectra for this object from AKARI or any other source prevents a direct comparison to the JWST NIRSpec spectra.

\par $\epsilon$ Ind Ba+Bb, has the second lowest chi-squared, and is shown in green in Figure \ref{fig:akari}. $\epsilon$ Ind Ba+Bb was discovered by \cite{scholz2003} as a companion to the K5V star, $\epsilon$ Ind A. \cite{mccaughrean2004} found that the object was actually a binary with spectral types T1 and T6.  \cite{yamamura2010} shows that the AKARI spectra exhibits the major molecular bands of CH$_4$, H$_2$O, CO, and CO$_2$. $\epsilon$ Ind Ba+Bb does not have overlapping near-IR spectral coverage with the SPHERE data, so we did not conduct further analysis.  Further, its status as a blended binary source complicates direct comparison to our dataset. 

\par From this empirical comparison, we are able to validate our identified spectral features based on the matching features in the AKARI spectral data and we are able to confirm the spectral type of HD 19467 B to be a late T dwarf. \citep{yamamura2010,sorahana2012}. Future brown dwarf observations with JWST NIRSpec will provide a more robust comparison when identifying individual molecular lines. 

\section{Forward Modeling with Atmospheric Grid Models and vertical mixing }\label{forward_modeling}

\subsection{Context: Vertical Mixing in planets and  cool sub-stellar objects}

Disequilibrium chemistry in Jupiter and Saturn has been extensively discussed in the literature (see \cite{2011ApJ...738...72V} and references therein). Vertical mixing has also been invoked to explain the atmospheric features of hot-Jupiters \citep{2011ApJ...737...15M}.  Such physical processes have also been observed in the cool T-type brown dwarfs \citep{1996ApJ...472L..37F,1997ApJ...489L..87N, 2014ApJ...797...41Z}. This vast body of work highlight the importance of the 3-5 $\mu$m range for characterizing atmospheric structures of exoplanets and substellar objects in the $\sim$ 500-1000 K regime. In particular early work by \cite{2004AJ....127.3516G} established an empirical sequence based on L and M band photometry. More recently \cite{2020AJ....160...63M} presented R $\sim$ 300 spectroscopic sequence  of seven brown dwarfs with effective temperatures between 750 K and 250 K in this wavelength range ($M$-band). All objects exhibit disequilibrium chemistry, with eddy diffusion coefficients $K_{zz}$, commensurate with vertical mixing, decreasing as effective temperature increases. Our NIRSpec observations of the T dwarf, HD 19467 B, provide a unique opportunity to repeat this experiment on a richer, higher resolution, and broader wavelength coverage dataset. This is what we carry out in this section. First, we use public forward model grids computed under the hypothesis of equilibrium chemistry to test how strongly our observation support disequilibrium chemistry. As a sanity check, we also conduct similar comparisons with published $J$- and $H$-band data and our high-pass filtered NIRSpec data, in case residual speckles still contaminate the continuum. Once we establish that no model with equilibrium chemistry can explain our data, we turn to new PHOENIX models that include effects of vertical mixing. \cite{2022ApJ...938..107M} presented grids exploring physical process and made precise prediction of how they would impact the JWST emergent spectrum of sub-stellar objects. However, since HD 19467 B lies at the hotter edge of \cite{2022ApJ...938..107M}’s grid, we leave comparisons with our data to further investigations.


\subsection{Models description and fitting method}

We forward model the data first using three grids, BTSETTL08 \citep{allard2012}, DRIFT-PHOENIX \citep{witte2011}, and Sonora Bobcat \citep{marley2021}. These grids use radiative-convective equilibrium but treat clouds differently and we chose these grids to explore how they would fit our new mid-IR moderate resolution spectra.  
The BTSETTL08 grid treats clouds with size distribution and number density as a function of depth based on nucleation, gravitational settling, and vertical mixing \citep{allard2012}. The DRIFT-PHOENIX grid treats clouds by modeling the effects of nucleation, surface growth, surface evaporation, gravitational settling, convection, and element conservation \citep{witte2011}. The Sonora Bobcat model grid is a cloudless model grid \citep{marley2021}. The model grids' effective temperature and surface gravity ranges used are in Table \ref{tab:param_ranges}. While we did not explore fitting metallicity and kept the grids at solar, different metallicities were not available for BTSETTL08 and NEWERA-PHOENIX, but metallicity ranges for Sonora bobcat are -0.5 to 0.5 and ranges for DRIFT-PHOENIX are -0.6 to 0.3. With the variety of cloud treatments in these grids, we are able to explore how they fit HD 19467 B's spectra in the 3-5$\mu$m range.

We fit the data using the prescription described in \cite{blake2010}, \cite{burgasser2016}, \cite{hsu2021}, and \cite{theissen2022} to determine the best-fit model from each grid. The effective temperature ($T_\mathrm{eff}$) and surface gravity ($\log g$) are estimated using a Markov Chain Monte Carlo (MCMC) method built using the \texttt{emcee} package which uses an implementation of the affine-invariant ensemble sampler \citep{goodman2010,foreman-mackey2013}. Our MCMC runs used 100 walkers, 500 steps, and a burn-in of 400 steps to ensure parameters were well-mixed. The details of the MCMC calculations follow those described in \cite{wilcomb2020}. The equations used for this routine are in the Appendix. We separately fit both the NIRSpec continuum and continuum-subtracted data because the continuum could be impacted by the speckle removal. We also fit the SPHERE data from \cite{mesa2020} to compare our best-fit values from in the near-infrared. Our forward modeling framework does include RV fits, but we do not quote these because we are uncertain about the wavelength solution for NIRSpec. An RV measurement would be an excellent measurement for future work. 

The output posterior distributions from the MCMC fits, a.k.a. corner plots, are included for reference in Appendix \ref{sec:app}.

\begin{figure*}
\centering
  \includegraphics[width=7in]{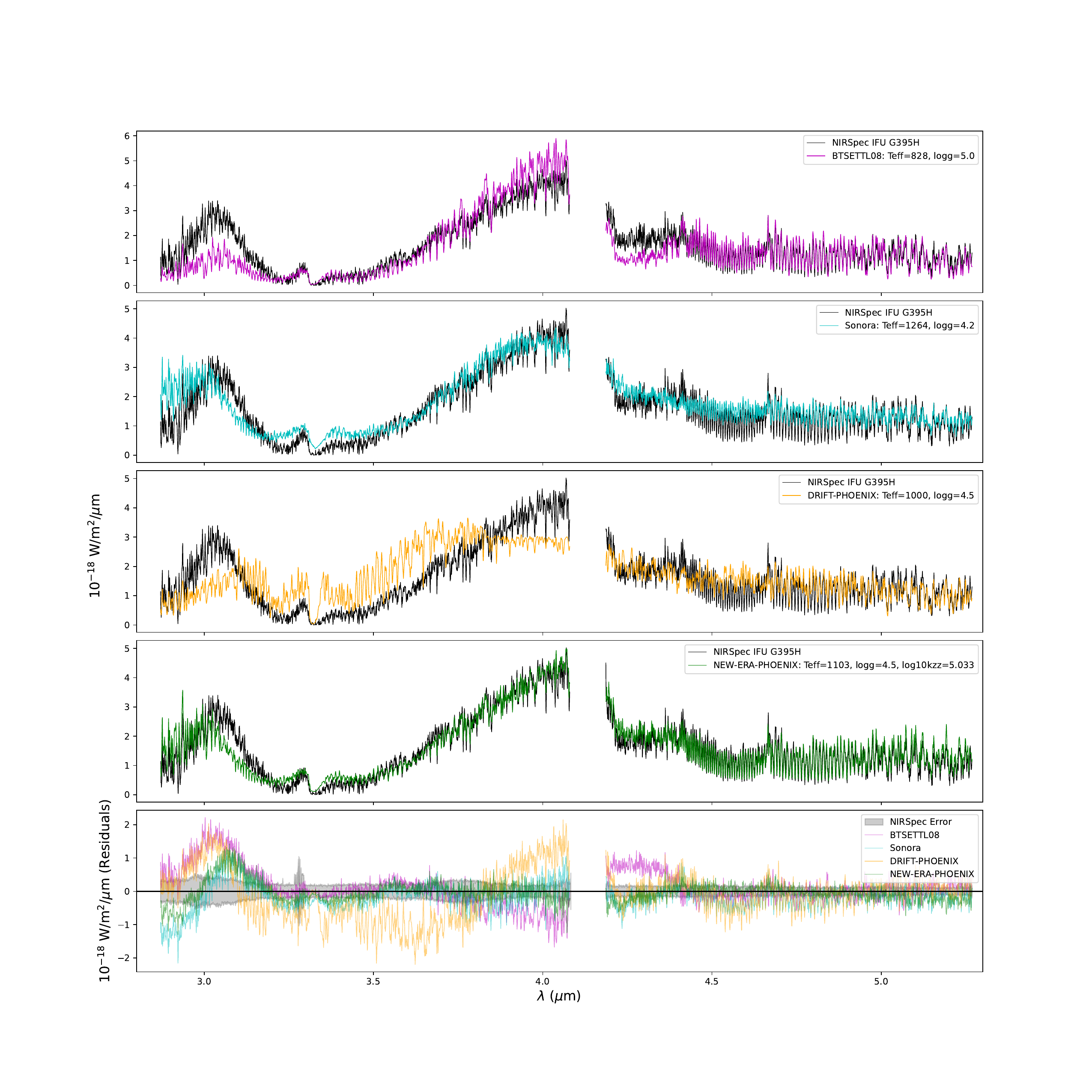}
\caption{The NIRSpec spectrum of HD 19467 B in black plotted against the best-fit models; BTSETTL08 in magenta in the top panel, Sonora Bobcat in blue in the second panel, DRIFT-PHOENIX in orange in the third panel, and NEWERA-PHOENIX, in green in the fourth panel. The residuals are plotted in the final panel in the corresponding colors to their models. A black line was drawn at zero, and the smallest residuals were from the NEWERA-PHOENIX model grid with the NIRSpec continuum errors plotted in gray.. This illustrates the need for updated models that include disequilibrium effects to properly fit the 3--5$\mu$m wavelength regime.}
\label{fig:continuum_bestfit}
\end{figure*}

\subsection{BTSETTL08}
We first fit the NIRSpec data including the continuum for HD 19467 B. We find a best-fit effective temperature of $T_\mathrm{eff}$=828$^{+4}_{-4}$ K and surface gravity $\log g$=5.0$^{+0.0}_{-0.001}$.  The best-fit is shown in the top panel of Figure \ref{fig:continuum_bestfit}.  The BTSETTL08 model fits the longer wavelength data very well except for the CO$_2$ feature at 4.2 $\mu$m, but under- and over-fits the continuum on either side of the 3.3 $\mu$m CH$_4$ dip. The fit also runs up against the edge of the grid in $\log g$ space, as shown in Figure \ref{fig:btsettl08_tri} in the appendix.



Next, we flatten the NIRSpec data and the models by high pass filtering them using \texttt{scipy.ndimage.filters.uniform$\_$filter} with a kernel size of 80 pixels.  We refit the data with the same set of BTSETTL08 models and find a best-fit effective temperature of $T_\mathrm{eff}=1100^{+2}_{-2}$ K and surface gravity $\log g=5.5^{+0.005}_{-0.008}$. The spectral fits are shown in the top panel of Figure \ref{fig:flat_bestfit} with the posterior distributions in the appendix in Figure \ref{fig:btsettl08_tri_flat}. 

We also fit the SPHERE IRDIS Long Slit Spectroscopy Medium Resolution Spectroscopy (LSS MRS)  spectra from \cite{mesa2020}. We find a best-fit effective temperature of $T_\mathrm{eff}$=905$^{+2}_{-4}$ K and surface gravity $\log g$=4.5$^{+0.016}_{-0.084}$ when forward modeling the spectra. These values are in between the values from using this grid to fit our NIRSpec spectra with and without the continuum. 

We find the effective temperature to be higher with the continuum subtracted spectra, but this has been noted and seen in previous work done with the OSIRIS IFU on the W. M. Keck Telescope \citep{wilcomb2020,hoch2022,hoch2023}. The surface gravities are in good agreement with previous values from the literature \citep{mesa2020,greenbaum2023}.



\subsection{DRIFT-PHOENIX}

With this grid we note that the temperature does not go below 1000 K, but we decided to explore using this grid to compare to our other fits. We find a best-fit effective temperature of $T_\mathrm{eff}$=1000$^{+0}_{-0}$ K and surface gravity $\log g$=4.5$^{+0.005}_{-0.008}$ shown in the third panel of Figure \ref{fig:continuum_bestfit} with the posterior distributions in Figure \ref{fig:drift_tri} in the appendix. We note that the temperature hits the edge of the grid, which was expected, with a slightly lower surface gravity than the BTSETTL08 grid fits.


The DRIFT-PHOENIX grid does not do a good job of fitting the CH$_4$ feature, or the CO forest. This could be due to the cloud modeling in the grid not representing the clouds in the cool T dwarf atmosphere. 

When forward modeling the high-pass-filtered NIRSpec spectra using the DRIFT-PHOENIX model grid, we find a best fit effective temperature of $T_\mathrm{eff}$=1212$^{+9}_{-8}$ K and surface gravity $\log g$=5.7$^{+0.03}_{-0.03}$. See the third panel in Figure \ref{fig:flat_bestfit} with the posteriors in Figure \ref{fig:drift_flat_tri} in the appendix. 

We also use this grid to fit the SPHERE LSS MRS low resolution spectra from \cite{mesa2020}. We find a best-fit effective temperature of $T_\mathrm{eff}$=2201$^{+3}_{-1}$ K and surface gravity $\log g$=6.0$^{+0.00}_{-0.007}$ when forward modeling the spectra. We note that these values are not close to previous values, with the surface gravity hitting the edge of the grid, or to the values obtained when fitting the NIRspec continuum and continuum-subtracted data. The grid is unable to fit the continuum of the low-resolution SPHERE spectra, missing the majority of the spectral features. We quote the resultant values here, but we do not include these in further analysis.




\subsection{Sonora Bobcat}
We find a best-fit effective temperature of $T_\mathrm{eff}$=1264$^{+6}_{-6}$ K and surface gravity $\log g$=4.2$^{+0.057}_{-0.058}$ when forward modeling the continuum NIRSpec spectra of HD 19467 B shown in the second panel of Figure \ref{fig:continuum_bestfit}. This model fits the continuum well on the redder end of the 3.3 $\mu$m CH$_4$ dip, but does not fit the CO$_2$ and CO features past 4 $\mu$m. The posterior distributions are shown in Figure \ref{fig:sonora_tri} in the appendix.



We flatten the NIRSpec data and the models and redo our forward modeling technique with the same set of Sonora models and find a best-fit effective temperature of $T_\mathrm{eff}$=1333$^{+12}_{-13}$ K and surface gravity $\log g$=4.9$^{+0.055}_{-0.060}$ shown in the second panel of Figure \ref{fig:flat_bestfit} with the posterior distributions shown in Figure \ref{fig:triangle_sonora_flat} in the appendix. 

Next, we fit the SPHERE spectra from \cite{mesa2020} using Sonora Bobcat. We find a best-fit effective temperature of $T_\mathrm{eff}$=1017$^{+4}_{-4}$ K and surface gravity $\log g$=3.5$^{+0.055}_{-0.034}$. 

Except for the flattened spectra surface gravity, Sonora Bobcat tended to give a higher temperature and lower surface gravity. This could be due to the treatment of non-equilibrium chemistry.

\begin{figure*}
\centering
  \includegraphics[width=7in]{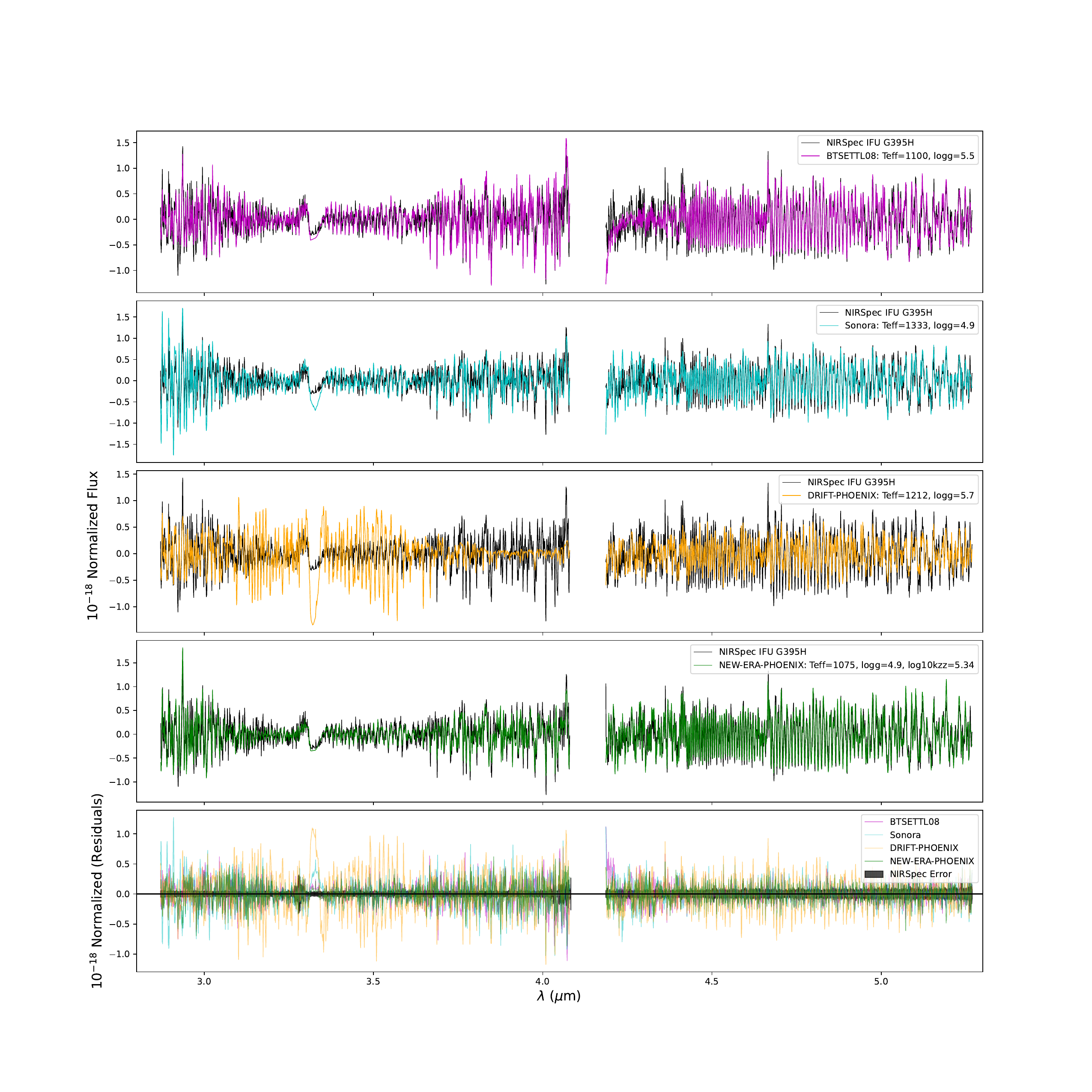}
\caption{Our continuum subtracted NIRSpec spectrum of HD 19467 B in black plotted against the best-fit models; BTSETTL08 in magenta in the top panel, Sonora Bobcat in blue in the second panel, DRIFT-PHOENIX in orange in the third panel, and NEWERA-PHOENIX, in green in the fourth panel. The residuals are plotted in the final panel in the corresponding colors to their models with the NIRSpec high pass filtered errors plotted in gray. A black line was drawn at zero, and the smallest residuals were from the NEWERA-PHOENIX model grid.}
\label{fig:flat_bestfit}
\end{figure*}





\begin{deluxetable*}{lccc}[!ht] 
\tabletypesize{\scriptsize} 
\tablewidth{0pt} 
\tablecaption{Summary of atmospheric parameters derived from MCMC fits.\label{tab:fits}}
\label{tab:atm_param}
\tablehead{ 
  \colhead{Spectra} & \colhead{Effective Temperature} & \colhead{Surface Gravity} & \colhead{Vertical Mixing} \\
  \colhead{HD 19467 B} & \colhead{${T}_\mathrm{{eff}}$ (K)} & \colhead{$\log g$} & \colhead{log$_{10}$($K_{zz}$)} 
}
\startdata 
\multicolumn{3}{c}{BTSETTL08} \\
\hline
NIRSpec Including Continuum & $828^{+4}_{-4}$ & $5.0^{+0.0}_{-0.001}$ & n/a \\
NIRSpec Continuum Subtracted & $1100^{+2}_{-2}$ & $5.5^{+0.005}_{-0.008}$ & n/a \\
SPHERE LSS MRS & $905^{+2}_{-4}$ & $4.5^{+0.016}_{-0.084}$ & n/a  \\
\hline
\multicolumn{3}{c}{DRIFT-PHOENIX} \\
\hline
NIRSpec Including Continuum & 1000$^{+0}_{-0}$ & 4.5$^{+0.005}_{-0.008}$ & n/a \\
NIRSpec Continuum Subtracted & 1212$^{+9}_{-8}$ & 5.7$^{+0.03}_{-0.03}$ & n/a \\
SPHERE LSS MRS & $2200^{+3}_{-1}$ & $6.0^{+0.001}_{-0.007}$ & n/a \\
\hline
\multicolumn{3}{c}{Sonora Bobcat} \\
\hline
NIRSpec Including Continuum & $1264^{+6}_{-6}$ & 4.2$^{+0.057}_{-0.058}$  & n/a \\
NIRSpec Continuum Subtracted & $1333^{+12}_{-13}$ & 4.9$^{+0.055}_{-0.060}$  & n/a   \\
SPHERE LSS MRS & 1017$^{+4}_{-4}$ & 3.5$^{+0.055}_{-0.034}$ & n/a \\
\hline
\multicolumn{3}{c}{NEWERA-PHOENIX} \\
\hline
NIRSpec Including Continuum & $1103^{+3}_{-2}$ & 4.5$^{+0.014}_{-0.011}$  & 5.03$^{+0.040}_{-0.037}$ \\
NIRSpec Continuum Subtracted & $1075^{+6}_{-7}$ & 4.9$^{+0.033}_{-0.032}$  & 5.34$^{+0.088}_{-0.088}$  \\
SPHERE LSS MRS & $1001^{+5}_{-3}$ & 4.0$^{+0.051}_{-0.028}$ & 9.069$^{+0.15}_{-0.16}$ \\
\hline
Adopted Values & 1103 & 4.5 & 5.03 \\
Allowed Range of Values & 1000--1200 & 4.1 - 5.0 & 5.0 - 5.4 \\
\enddata
\end{deluxetable*}

\begin{figure*}
\centering
  \includegraphics[width=7in]{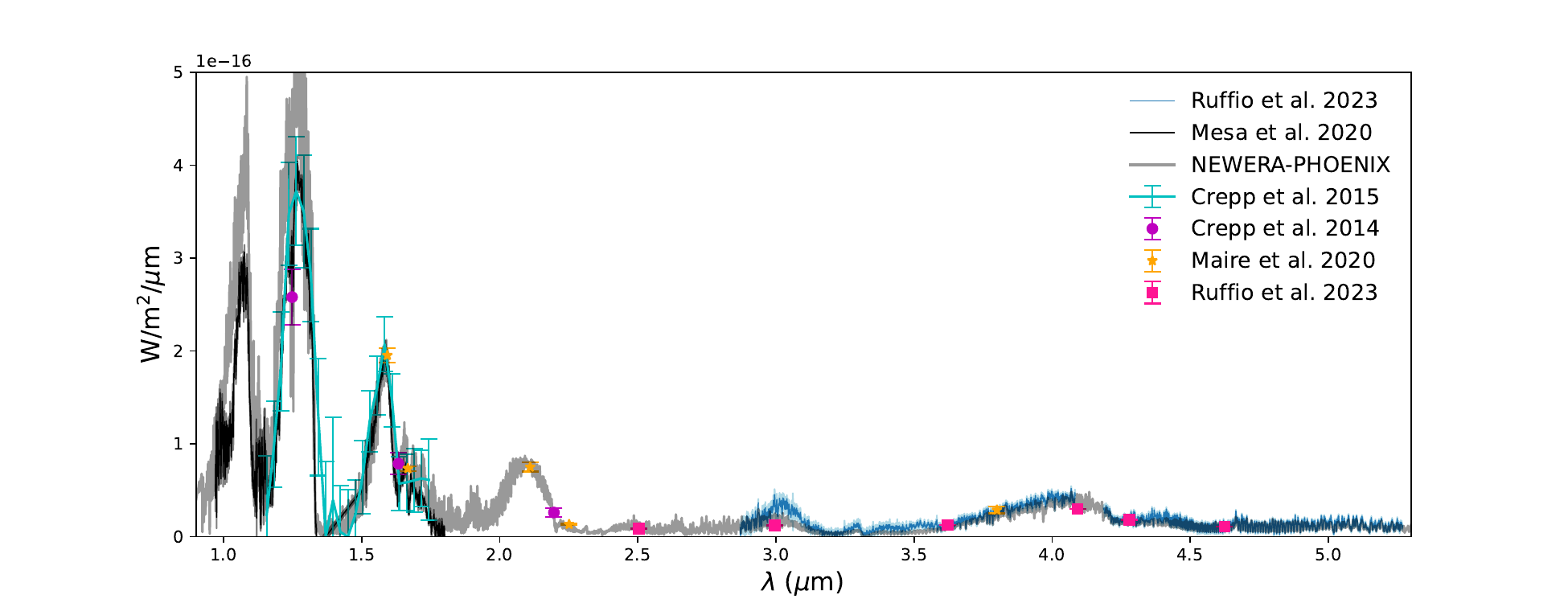}
\caption{Compilation of all literature photometry and spectroscopy on HD 19467 B, including JWST NIRCam coronagraph data published in Greenbaum et al. and re-reduced using SpaceKLIP and historical near IR ground based data. The near IR spectra from \cite{mesa2020} and \cite{crepp2015} exhibit broad deep absorption features that are characteristics of T dwarfs but they lack the sensitivity and spectral resolution to pick up the less abundant molecules identified with JWST. The best-fit model from the NEWERA-PHOENIX grid is overplotted in light gray behind all of the available data. The model provides an excellent fit for most photometric data, but in the bluer wavelengths it overestimates the flux slightly. This could be caused be a variety of reasons, including the speckle subtraction in that area of the spectrum as well as not including clouds in our NEWERA-PHOENIX grid.}
\label{fig:megaplot}
\end{figure*}

\subsection{NEWERA-PHOENIX Spectral Modeling}\label{custom_models}
\par Since none of the models assuming equilibrium chemisty are able to capture all the molecular features in our NIRSpec data,  we also use NEWERA-PHOENIX models from Barman et al. in prep and Hasuchildt et al. in prep that include vertical mixing inducing disequilibrium chemistry. The grid of synthetic spectra was calculated with the PHOENIX model atmosphere code \citep{hauschildt1999} adopting solar atomic abundances \citep{asplund2009}. The current grid spans ${T}_\mathrm{{eff}}$ = 800--1350 K and $\log g$ = 4--5.5, which spans the range of values from Section \ref{forward_modeling}, and is only a subset of the much larger NEWERA-PHOENIX grid that will be made publicly available. Chemical equilibrium is initially assumed for all species, including condensates. The models are in hydrostatic and radiative-convective equilibrium. To simulate cloud free conditions, the condensate opacities are not included in the radiative transfer calculation (similar to the “COND” models of \citealp{allard2001}). Mixing ratios of certain molecules (H$_2$O, CO, CO$_2$, CH$_4$, N$_2$, NH$_3$) are modified to account for disequilibrium brought about by vertical mixing, following the procedure in \cite{barman2011} and parameterized by the eddy diffusion coefficient ($K_{zz}$, with grid values of 10$^2$, 10$^4$, 10$^6$, 10$^8$). The molecular line data are continuously updated in PHOENIX and includes the latest recommended data from ExoMol \citep{tennyson2020}, HITRAN \citep{tan2022} and HITEMP \citep{hargreaves2020}. The line data for the molecules that have prominent absorption features across our NIRSpec data are from the following sources: H$_2$O \citep{polyansky2018}, CO \citep{hitemp2010}, CO$_2$ \citep{hitemp2010,li2015}, and CH$_4$ \citep{Yurchenko2014,Yurchenko2017}. The details about the latest line list selection process and the complete list of line data included with PHOENIX will be reported in future work in Hauschildt et al. in prep.  

We fit our NIRSpec continuum and continuum-subtracted spectra again, but this time we use the NEWERA-PHOENIX grid described above which includes the third parameter, $K_{zz}$. We find a best fit effective temperature of $T_\mathrm{eff}$=$1103^{+3}_{-2}$ K, a surface gravity of $\log g$=4.5$^{+0.014}_{-0.011}$, and a log$_{10}$($K_{zz}$)=5.03$^{+0.03}_{-0.037}$  shown in the fourth panel of Figure \ref{fig:continuum_bestfit}. This model fits the continuum best of all the grids on the redder end of the 3.3 $\mu$m CH$_4$ dip as well as the CO$_2$ and CO features passed 4 $\mu$m. The posterior distributions are shown in Figure \ref{fig:corner_new_era} in the appendix.

We then fit the flattened NIRSpec data using the high pass filtered NEWERA-PHOENIX models and find a best fit effective temperature of $T_\mathrm{eff}$=$1075^{+7}_{-7}$ K, a of surface gravity $\log g$=4.9$^{+0.029}_{-0.037}$, and a log$_{10}$($K_{zz}$)=5.344$^{+0.1}_{-0.084}$ shown in the fourth panel of Figure \ref{fig:flat_bestfit} with the posterior distributions shown in Figure \ref{fig:corner_new_era_flat} in the appendix. 

Next, we fit the SPHERE spectra from \cite{mesa2020} using the NEWERA-PHOENIX grid as well. We find a best fit effective temperature of $T_\mathrm{eff}$=1001$^{+5}_{-3}$ K, a surface gravity of $\log g$=4.0$^{+0.051}_{-0.028}$, and a log$_{10}$($K_{zz}$)=9.069$^{+0.153}_{-0.164}$ when forward modeling the spectra. The wavelength regime of the SPHERE data is not as sensitive to non-equilibrium chemistry, which could point towards why $K_{zz}$ is the one parameter that is not consistent with the rest of our results. 

\subsection{Summary of Forward Modeling}
When looking at the forward modeling fits for our NIRSpec G395H data, the continuum-subtracted results are similar to the continuum results, but are not consistent within the error bars, which can be expected. 


At the time we have no way to assess the impact of residual systematic errors from starlight noise. While we expect the method in \cite{Ruffio2023} to minimize them, we have not yet conducted injection recovery tests that are commonly used to assess the impact of residual speckles in high contrast IFS data (i.e., \cite{agrawal2023}). As a result, we also consider high-pass-filtered, continuum-subtracted spectra, that removes residual features at low spatial frequencies in the continuum of the spectra. 
We also do not take into account our DRIFT-PHOENIX results in our final reported value range due to the poor fits of the models with our NIRSpec data and the \cite{mesa2020} SPHERE spectra. We adopt values of $T_\mathrm{eff} = 1103$~K, with a range of 1000--1200~K, $\log g=4.5$, with a range of 4.14--5.0, and log$_{10}$($K_{zz}$)=5.03, with a range of 5.0--5.44 summarized in Table \ref{tab:atm_param}. The error bars from the fits in Table \ref{tab:atm_param} are the statistical error bars and those are too small to account for systematics. The ranges for our adopted values were chosen based on the best-fit values across the models to incorporate the systematic uncertainties that are not fit for properly by our forward modeling framework. In particular, we used the size of the grid steps in the NEWERA-PHOENIX model grid to encompass interpolation errors between steps. Our best-fit NEWERA-PHOENIX model is a reasonable fit for most of the data from the literature as well shown in Figure \ref{fig:megaplot}, where we broadened the lines of the model to fit the resolution of JWST's NIRSpec. While it is possible that lower temperatures could be invoked for HD 19467 B, this is one of the first times we have been able to model a companion atmosphere in the 3--5 $\mu$m range at the resolution of NIRSpec. Therefore, further modeling of brown dwarf and exoplanet atmospheres using NIRSpec will reveal how this wavelength range will impact previous results using low resolution and ground-based spectra, and will inform how to improve current atmosphere model grids.

\section{Molecular Analysis with PHOENIX Models}\label{molecules}
Having established bulk parameters, T$_\mathrm{eff} = 1103$~K, $\log g=4.5$, and established the degree of vertical mixing driving the observed disequilibrium chemistry, log$_{10}$($K_{zz}$)=5.03, we now explore the abundances of the molecules detected in our spectra. The best-fit parameters shown on Figure \ref{fig:continuum_bestfit} and Figure \ref{fig:flat_bestfit} were obtained using the atomic abundances are those of the solar composition from \cite{asplund2009}. Thus, our starting baseline abundances of Carbon and Oxygen are 8.43 and 8.69 respectively (we use a log scale where log A(H) is 12), with a C/O = 0.55 for the NEWERA-PHOENIX model that yields our best-fit bulk parameters. We aim to characterize the mixing ratios in this section to test disequilibrium chemistry in the atmosphere of HD 19467 B and see if the mixing ratios match those of Solar in the main NEWERA-PHOENIX grid. 

To test these mixing ratios, we create a smaller grid around the best-fit parameters, but now vary molecular abundances of H$_2$O, CO, CO$_2$, and CH$_4$. 
In this smaller grid,  we scale the mixing ratios of H$_2$O, CO, CO$_2$, and CH$_4$ by factors ranging from 0.1 to 10 one at a time by holding the rest of the mixing ratios at solar. The synthetic spectra were created for specific wavelength ranges in the NIRSpec data based on the molecular absorption features. Synthetic spectra that varied H$_2$O and CH$_4$ span the 2.9--4.0 $\mu$m range, and synthetic spectra that varied CH$_4$ and CO span the 4.2--5.2 $\mu$m range. 

\begin{figure*}[ht!]
\centering
  \includegraphics[width=3in]{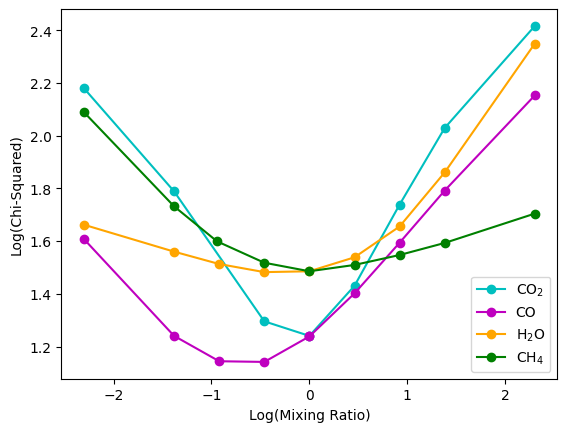}
\caption{Results of $T_\mathrm{eff}$ = 1103$^{+97}_{-103}$ K, $\log g$ = 4.5$^{+0.5}_{-0.36}$, and log$_10$($K_{zz}$) = 5.03$^{+0.414}_{-0.030}$ model fits with varying abundances for H$_2$O, CO, CO$_2$, and CH$_4$ to our continuum-subtracted NIRSpec spectrum. The abundances are given in units relative to the ratio in the Sun, such that a value of zero implies the solar value.  The scalings of most molecules prefer solar. CO could be slightly subsolar, but with the distributions being broad they are consistent with no scaling at all. For CO$_2$ and CH$_4$ the mixing ratios were 1.000$^{+0.585}_{-0.369}$. For CO and H$_2$O the mixing ratios were 0.631$^{+0.369}_{-0.233}$. These ratios are all consistent with a Solar C/O ratio, which is to be expected in a T5.5 brown dwarf.}
\label{fig:chisq}
\end{figure*}

Once we had synthetic spectra for each wavelength range in the NIRSpec spectra, we conducted a chi-sqaured analysis to determine the lowest chi-squared value for each of the models, focusing on one molecular mixing ratio at a time. To focus only on the line absorption, these scaled models were compared to the observations after the continuum was subtracted. 

For H$_2$O and CH$_4$, we took all of the models that varied the mixing ratio of H$_2$O while holding CH$_4$ at solar and calculated chi-squared values divided by the degrees of freedom for each mixing ratio. These are plotted in orange in Figure \ref{fig:chisq}. We do the same calculation for CH$_4$ and plot the values in green in Figure \ref{fig:chisq}. 

For CO$_2$ and CO, we used the redder part of the flattened spectrum from 4.2--5.3 $\mu$m and compared it to our flattened models that vary the mixing ratios of the two molecules. We did the same analysis as we did with H$_2$O and CH$_4$. These values for CO$_2$ are plotted in blue in Figure \ref{fig:chisq}. We conduct the same analysis on the models that vary CO and obtain the curve plotted in purple in Figure \ref{fig:chisq}.

Except for H$_2$O and CO, the chi-squared was minimized when the mixing ratio scale factor was 1.0. We calculate the 1 - $\sigma$ uncertainties by taking the values from models within $\pm$ 1 of our lowest chi-squared which were 1.000$^{+0.585}_{-0.369}$. For H$_2$O and CO, scaling the mixing ratios by 0.7 and 0.5, respectively, resulted in the lowest chi-squared; However, the chi-squared distributions are very broad, indicating that the scaling of the molecules did not have a large impact on the chi-sqaured values and is consistent with no scaling at all (Fig. \ref{fig:chisq}). For H$_2$O and CO, the best molecular scaling fraction was 0.631$^{+0.369}_{-0.233}$, which includes solar. These small differences are most likely a consequence of the treatment of vertical mixing and disequilibrium chemistry in the NEWERA-PHOENIX models. Since the values are consistent with solar, we conclude that the C/O ratio of the companion is consistent with the solar value as well. \cite{maire2020} also measured a C/O of the host star to be 0.52, which is also consistent with the value of the Sun. It is not surprising that HD 19467 B has a similar C/O value to its host star, as it most likely formed from the same material as its host due to its large mass \citep{oberg2011}.

\section{Dynamical Mass of HD~19467~B}\label{dyn_mass}

In order to refine the dynamical mass of HD~19467~B, we perform an orbit fit that combines radial velocities, relative astrometry, and astrometric accelerations using \texttt{orvara} as described in \citet{brandtorvara2021}. 

The radial velocities are taken from both the re-calibrated HARPS data \citep{trifonov2020} and HIRES radial velocities, where the latest reduction is described in \citet{rosenthal2021}, with additional new HIRES radial velocities as reported in \citet{greenbaum2023}. The relative astrometry values come from multiple imaging epochs \citep{crepp2014, maire2020, bowler2020astrometry}, with the full list of relative astrometry used for the orbit fit listed in Table~\ref{tab:relative-astrometry}.

In addition, we derive updated relative astrometry values from the HD~19467~B NIRCam observations \citep{greenbaum2023} using SpaceKLIP as described in \citet{Kammerer2022}, with the new values listed in Table~\ref{tab:relative-astrometry}. For the NIRCam relative astrometry, we incorporate a systematic error of 7~mas in quadrature to account for the uncertainty in determining the position of the star behind the coronagraphic mask, as noted in \cite{greenbaum2023}, where for the orbital fit we use an averaged astrometric point from the updated JWST/NIRCam values presented in Table~\ref{tab:relative-astrometry}. The newly derived relative astrometry values are in agreement with the values from \citet{greenbaum2023} within the uncertainties with minor differences between the values. We do not derive any relative astrometric values from the NIRSpec data due to not being able to accurately determine the position of the host star due to the saturation of the star on the detector.

For the orbit fit, we use \texttt{orvara} \citep{brandtorvara2021} which is specifically designed to combine radial velocities, relative astrometry, and absolute astrometry in order to derive precise dynamical masses as demonstrated recently in \cite[e.g.,][]{brandt2021,rickman2022,rickman2024}. The \texttt{orvara} package makes use of astrometric data from the Hipparcos-Gaia Catalog of Accelerations \citep[HGCA,][]{brandt2018, brandtHGCA2021} that provide a long baseline of data between the astrometric measurements taken with Hipparcos \citep{esa1997, vanleeuwen2007, Hipparcos..JAVA} and from \emph{Gaia} eDR3 \citep{gaia2021}.

\begin{table*}[]
    \centering
    \begin{tabular}{cccccc}
    \hline
    \hline
       Epoch & Instrument & Filter & Separation ('') & Position Angle (deg) & Reference \\
    \hline
        2011.66 & Keck/NIRC2 & $K'$ & $1.6627\pm0.0049$ & $243.14\pm0.19$ & \citet{crepp2014} \\
        2012.02 & Keck/NIRC2 & $H$ & $1.6657\pm0.0070$ &	$242.25\pm0.26$ & \citet{crepp2014} \\ 
        2012.02 & Keck/NIRC2 & $K'$ & $1.6573\pm0.0072$ & $242.39\pm0.38$ & \citet{crepp2014} \\
        2012.66 & Keck/NIRC2 & $K'$ & $1.6618\pm0.0044$ & $242.19\pm0.15$ & \citet{crepp2014} \\
        2012.76 & Keck/NIRC2 & $K_s$ & $1.6531\pm0.0041$ & $242.13\pm0.14$ & \citet{crepp2014} \\ 
        2017.76 & VLT/NACO & $L'$ & $1.6370\pm0.0190$ & $238.68\pm0.47$ & \citet{maire2020} \\
        2017.84 & VLT/SPHERE & $K1$ & $1.6367\pm0.0018$ & $239.39\pm0.13$ & \citet{maire2020} \\
        2017.84 & VLT/SPHERE & $K2$ & $1.6344\pm0.0050$ & $239.44\pm0.21$ & \citet{maire2020} \\
        2018.80 & VLT/SPHERE & $H2$ & $1.6314\pm0.0016$ & $238.88\pm0.12$ & \citet{maire2020} \\
        2018.80	& VLT/SPHERE & $H3$ & $1.6314\pm0.0016$ & $238.88\pm0.12$ & \citet{maire2020} \\
        2018.80	& Keck/NIRC2 & $H$ & $1.628\pm0.005$ & $239.50\pm0.30$ &  \citet{bowler2020astrometry} \\
        2022.62 & JWST/NIRCam & F250M & $1.607\pm0.0110$  & $235.47\pm0.19$ & This work \\ 
        2022.62 & JWST/NIRCam & F300M & $1.618\pm0.0074$  & $236.55\pm0.05$ & This work \\
        2022.62 & JWST/NIRCam & F360M & $1.606\pm0.0071$  & $236.89\pm0.04$ & This work \\
        2022.62 & JWST/NIRCam &	F410M & $1.602\pm0.0071$  & $236.85\pm0.04$ & This work \\
        2022.62	& JWST/NIRCam &	F430M & $1.613\pm0.0071$  & $236.94\pm0.04$ & This work \\
        2022.62 & JWST/NIRCam &	F460M & $1.607\pm0.0079$  & $236.80\pm0.08$ & This work \\
    \hline
    \end{tabular}
    \caption{Relative astrometry of HD~19467~B.}
    \label{tab:relative-astrometry}
\end{table*}

\begin{table*}[]
    \centering
    \begin{tabular}{cccc}
    \hline
    \hline
       Parameters & Units & Prior & HD~19467~B \\
    \hline
    Fitted Parameters \\
    \hline
       Companion mass $M_{\rm{comp}}$ & $M_{\rm{Jup}}$ & $1/M_{\mathrm{comp}}$ & ${71.6}_{-4.6}^{+5.3}$ \\
       Host-star mass $M_{\rm{host}}$ & $M_{\rm{\odot}}$ & $\mathcal{N}(0.95, 0.02)$ & $0.95\pm0.020$ \\
       Parallax & mas & $1/\pi$ & $31.22\pm0.02$ \\
       Semimajor axis $a$ & AU & $1/a$ & ${46.9}_{-7.4}^{+11.0}$ \\
       Inclination $i$ & $\degree$ & $\mathcal{U}(0, 180)$ & ${134.7}_{-6.9}^{+12.0}$ \\
       Jitter & $\sigma$ & $1/\sigma$ & ${3.13}_{-0.22}^{+0.25}$ \\
       $\sqrt{e}\sin\omega$ & & $\mathcal{U}(-1, 1)$ & ${-0.67}_{-0.07}^{+0.05}$ \\
       $\sqrt{e}\cos\omega$ & & $\mathcal{U}(-1, 1)$ &${0.12}_{-0.22}^{+0.17}$ \\
       PA of the ascending node $\Omega$ & $\degree$ & $\mathcal{U}(-180, 180)$ & ${39.1}_{-16}^{+8.5}$ \\
       \hline
       Derived Parameters \\
       \hline
       Orbital period $P$ & yr & & ${319}_{-72}^{+114}$ \\
       Eccentricity $e$ & & & ${0.50}_{-0.08}^{+0.10}$\\
       Time of periastron $T_0$ & JD & &  ${2486750}_{-1980}^{+1523}$ \\
       Semimajor axis & mas & & ${1465}_{-230}^{+332}$ \\
       Argument of periastron $\omega$ & $\degree$ & & ${280}_{-18}^{+14}$ \\
       Mass ratio $q$ & $M_{\rm{comp}}/M_{\rm{host}}$ & & $0.072^{+0.006}_{-0.005}$ \\
    \hline
    \end{tabular}
    \caption{Orbital parameters and dynamical mass of HD~19467~B.}
    \label{tab:orbital-parameters}
\end{table*}

\begin{figure*}[htp!]
    \centering
    \includegraphics[width=0.50\textwidth]{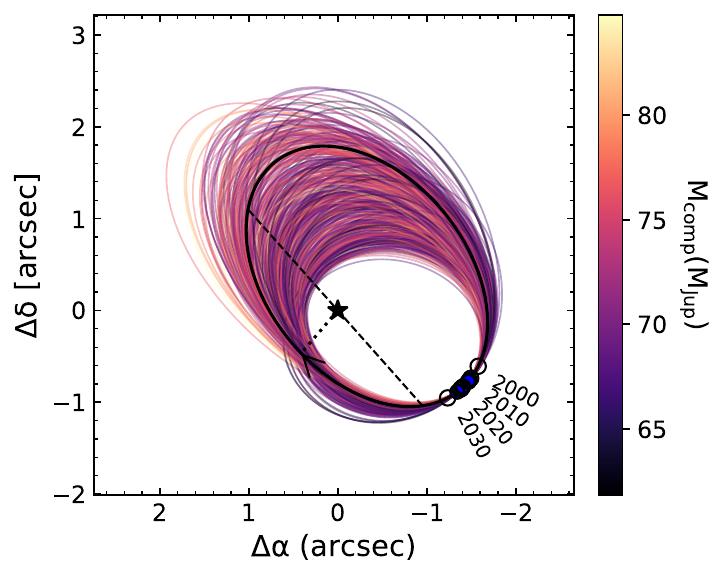}
    \includegraphics[width=0.45\textwidth]{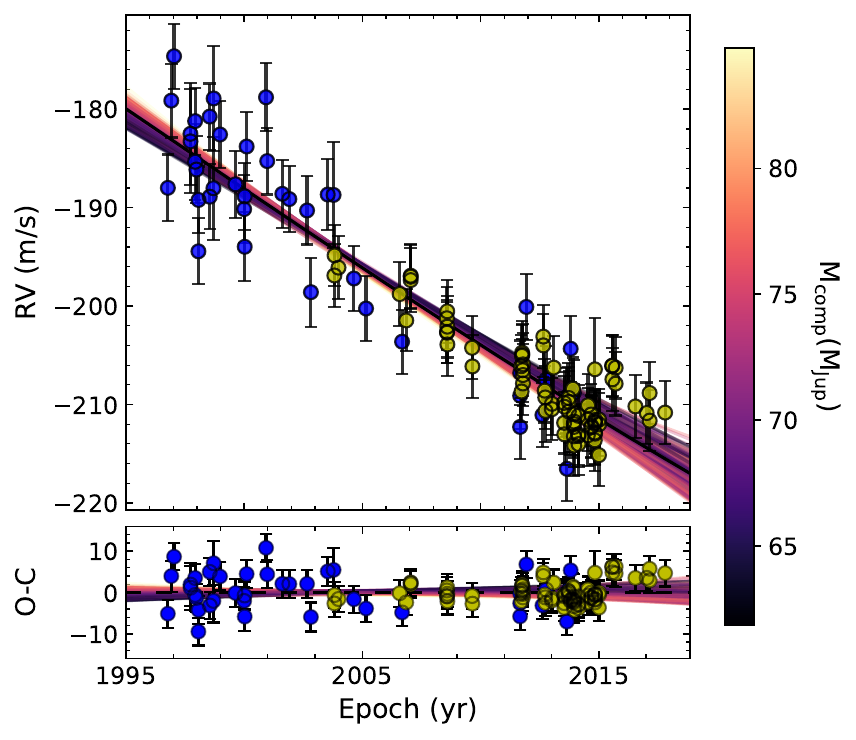}
    \caption{\textbf{Left:} Orbit fit of HD~19467~B using the orbit-fitting code \texttt{orvara} \citep{brandtorvara2021}. The thick black line represents the highest likelihood orbit; the thin colored lines represent 500 orbits drawn randomly from the posterior distribution. Dark purple corresponds to a low companion mass and light yellow corresponds to a high companion mass. The dotted black line shows the periastron passage, and the arrow at the periastron passage shows the direction of the orbit. The dashed line indicates the line of nodes. Predicted past and future relative astrometric points are shown by black circles with their respective years, while the observed relative astrometric points are shown by the blue-filled data points, where the measurement error is smaller than the plotted symbol. \textbf{Right:} Radial-velocity changes induced by HD~19467~B. Shown are the radial-velocity data from HIRES (blue points) and HARPS (yellow points). Again the thick line shows the highest likelihood fit; the thin colored lines show 500 orbits drawn randomly from the posterior distribution. The residuals of the radial velocities are shown in the bottom panel.}
    \label{fig:orbit_fit}
\end{figure*}

\begin{figure*}[hbp!]
    \centering
    \includegraphics[width=0.46\textwidth]{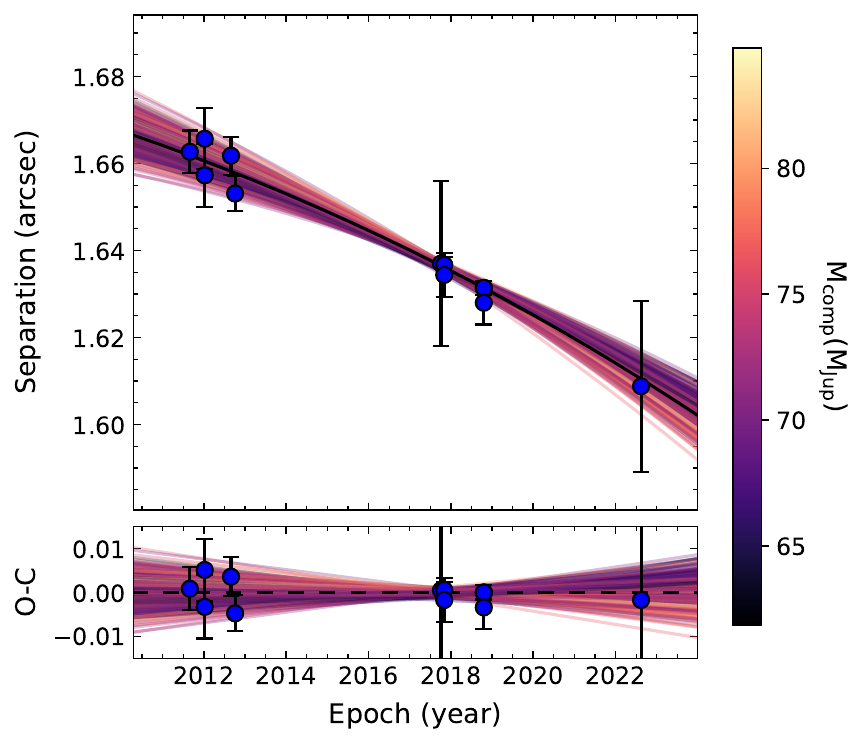}
    \includegraphics[width=0.45\textwidth]{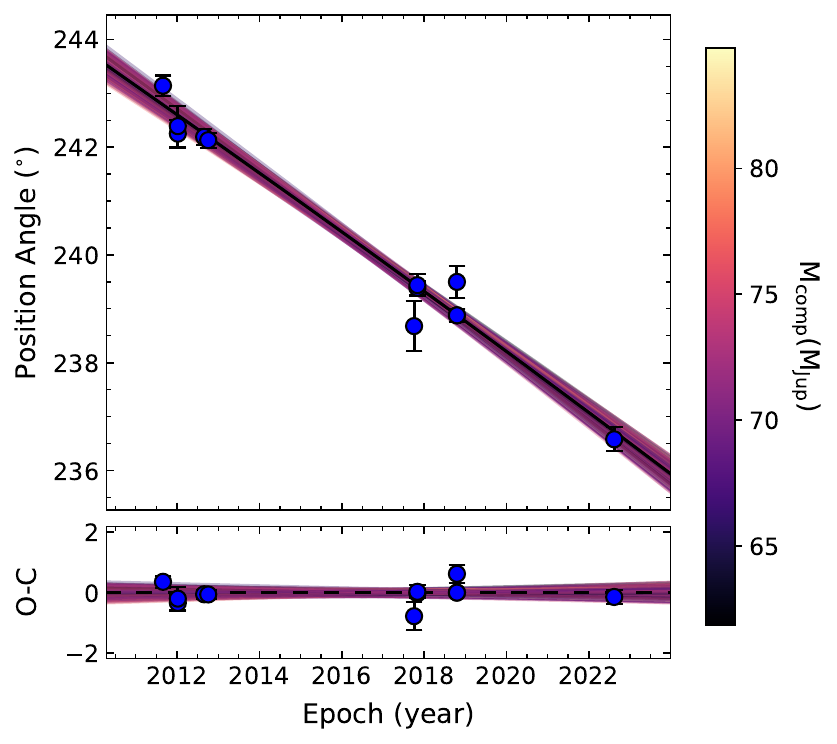}
    \caption{Relative separation (left) and position angle (right) of HD~19467~B across multi epochs of relative astrometry as listed in Table~\ref{tab:relative-astrometry}. The residuals are shown in the bottom two panels.}
    \label{fig:relative-astrometry}
\end{figure*}

\begin{figure*}
    \centering
    \includegraphics[width=0.95\textwidth]{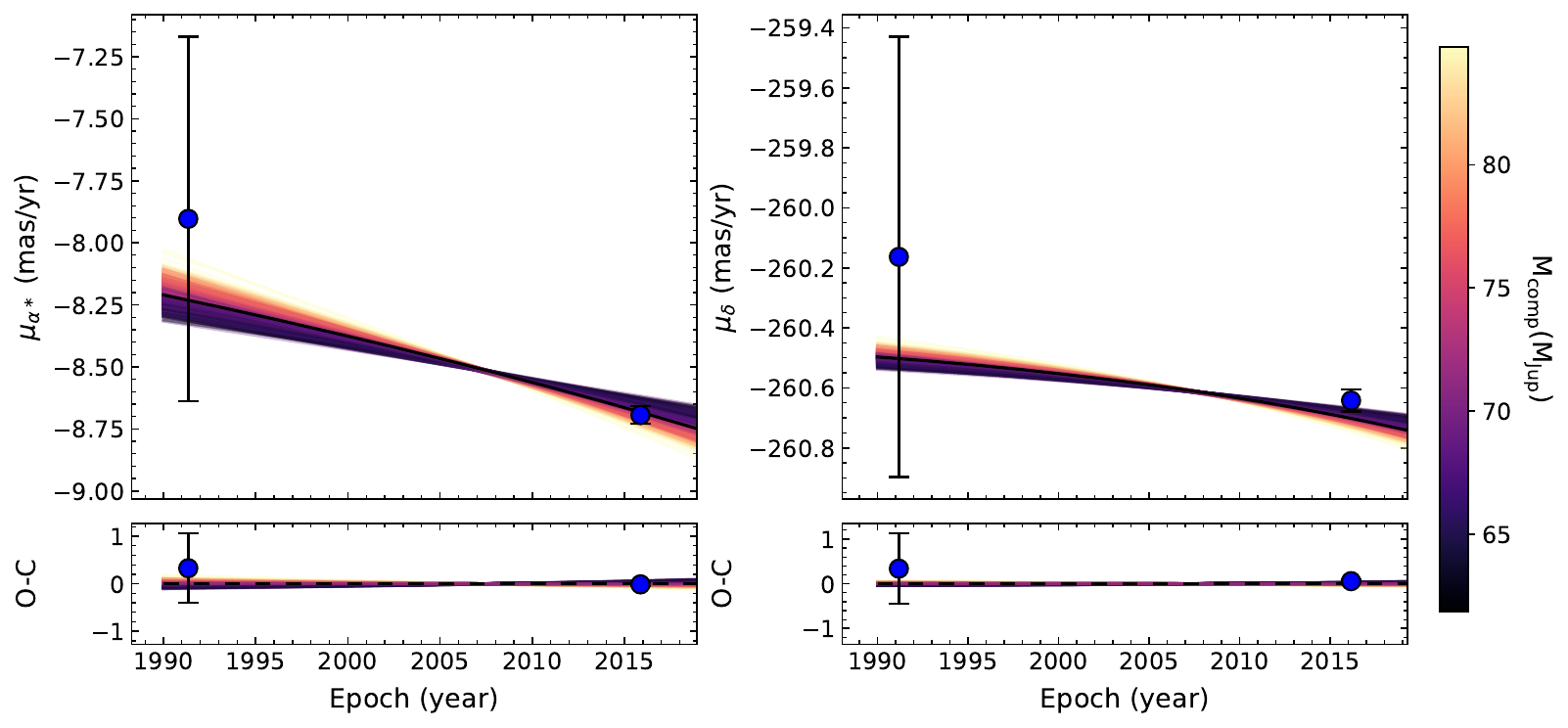}
    \caption{Acceleration induced by the companion on the host star as measured from absolute astrometry from HIPPARCOS and Gaia for $\mu_{\alpha}$ (left) and $\mu_{\delta}$ (right). The thick black line represents the highest likelihood orbit; the thin colored lines are 500 orbits drawn randomly from the posterior distribution. Darker purple represents a lower companion mass and light yellow represents a higher companion mass. The residuals of the proper motions are shown in the bottom panels.}
    \label{fig:proper-motions}
\end{figure*}

We adopt a prior of $0.95\pm0.02M_{\odot}$ on the mass of the primary star, as the same in \citet{brandt2021} and \citet{maire2020} and imposed uninformative log-flat prior on the mass of HD~19467~B, and uniform priors on the remaining orbital parameters. We run \texttt{orvara} with 30,000 steps in each chain. The resulting orbital parameters are listed in Table~\ref{tab:orbital-parameters}.

For the orbit fits shown in Fig.~\ref{fig:orbit_fit}, we implemented a burn-in phase of 300 multiplied by every 50th step that is saved on the chain and randomly drew 500 orbits showing the posterior distribution. The highest likelihood fit for the RVs is also shown in Fig.~\ref{fig:orbit_fit}, in addition to the fits to relative separation and position are shown in Fig.~\ref{fig:relative-astrometry}, and the fits to the proper motions from Hipparcos and \emph{Gaia} are shown in Fig.~\ref{fig:proper-motions}. The posterior distributions of the orbital parameters are included in the Appendix in Fig.~\ref{fig:orbit-corner-plot}.

From the orbital fit, we find the mass of HD~19467~B to be $71.6^{+5.3}_{-4.6}~M_{\rm{Jup}}$. Our updated mass estimate for HD~19467~B is within 1~$\sigma$ of the estimates of $74^{+12}_{-9}~M_{\rm{Jup}}$ from \cite{maire2020}; $65.4^{+5.9}_{-4.6}~M_{\rm{Jup}}$ from \cite{brandt2021}, and $81^{+14}_{-12}~M_{\rm{Jup}}$ from \cite{greenbaum2023}, although our mass estimate seems to favor the lower end of the mass derived in \cite{greenbaum2023}. Both our dynamical mass estimate and orbital parameters including the eccentricity and orbital period are within $1~\sigma$ agreement with previously derived values from \cite{maire2020}, \cite{brandt2021}, and \cite{greenbaum2023}.

\section{Discussion}\label{discussion}

\par  In this work, we presented JWST NIRSpec's first high contrast imaging spectroscopic analysis of the benchmark brown dwarf, HD 19467 B. While this system is much older ($\sim$9 Gyr) than most directly imaged planetary systems, it is well studied from the ground, has a dynamical mass measurement, and, with a moderate contrast ($\sim$1e-6), provides a good test case for high contrast imaging spectroscopy using the NIRSpec IFU. With the reduction techniques described in the companion paper, \cite{Ruffio2023}, and the forward modeling results from this work, the groundwork has been forged to push the field of direct imaging spectroscopy into longer wavelengths and create new opportunities to understand their atmospheres.

\subsection{Implications for Atmospheric Models}

\par  To characterize the atmosphere of HD 19467 B, we forward modeled the continuum NIRSpec data and the continuum subtracted data using a variety of atmospheric grids. We chose to use four grids to explore how the different models, and their assumptions, would fit the new NIR data from JWST. As stated in Section \ref{forward_modeling}, we chose BTSETTL08 \citep{allard2012}, DRIFT-PHOENIX \citep{witte2011}, Sonora Bobcat \citep{marley2021}, and the NEWERA-PHOENIX grid with updated line lists from Hauschildt et al. in prep. These atmospheric model grids are  well known, and widely used throughout the exoplanet community. With the differing treatment of clouds and chemistry in the grids, we can see how well the assumptions hold in the NIR for the case of HD 19467 B.  

\par Our results show that BTSETTL08, Sonora Bobcat, and the NEWERA-PHOENIX grid are able to replicate the spectral features and continuum well, but DRIFT-PHOENIX was unable to do the same. This is most likely due to the microphysics cloud model used, indicating that the chemistry in HD 19467 B's atmosphere is not well represented with the processes included. (This is to be expected, since a T5 atmosphere should not have visible clouds in the photosphere \citep{burgasser2002}). BTSETTL08 was the best-fit out of the public grids fitting the CO features from about 4.3 $\mu$m and on, but over estimates the CO$_2$ feature and is unable to replicate the CH$_4$ spectral shape at 3.3 $\mu$m. BTSETTL08 also has a microphysics cloud model described above, but does not include non-equilibrium chemistry which could explain why it struggles to fit both the CH$_4$ feature and the CO$_2$ feature. Sonora Bobcat does a good job replicating the continuum and the CH$_4$ feature, but is unable to provide a good fit to the CO$_2$ and CO features in the longer wavelengths of our NIRSpec data. This could be caused by not including non-equilibrium chemistry effects. 

The NEWERA-PHOENIX model grid does the best overall for this object fitting the CH$_4$, CO$_2$, and CO the best simultaneously. This is because our new grid has modified mixing ratios of molecules to account for disequilibrium chemistry caused by vertical mixing in the atmosphere. The H$_2$O continuum feature in the shorter wavelengths is not well fit by any of the grids and could be caused by an artifact from our reduction methods, or it could be a missing assumption from the models. 

\subsection{Additional Atmospheric Information from 3--5 $\mu$m}

\par Before JWST, the best way to access high signal-to-noise spectra, due to the brightness of Earth's atmosphere, of substellar objects from $\sim$3--5$\mu$m was the Japanese AKARI satellite. \cite{yamamura2010} discovered that CO band strength in late-L to late-T dwarfs were not consistent with predictions and explained the discrepancy via inadequate modeling of vertical mixing effects. They also found that there was a larger abundance of CO$_2$ in three objects in their sample than expected, with atmospheric models either over or under predicting the abundance. A possible reason was postulated to be a higher than solar C and O elemental abundances used in previous studies \citep{tsuji2011}, which was also found by \cite{sorahana2014}. CO$_2$ is a ubiquitous feature in late-L to late-T type dwarf spectra, which was previously unknown without the longer wavelength range that space-based observing provides. 

\par JWST now provides access to substellar spectra with wavelengths beyond 3 $\mu$m with moderate resolution. \cite{greenbaum2023} conducted atmospheric modeling of HD 19467 B using NIRCam photometric points and the new Sonora Cholla grid. This grid includes effects of non-equilibrium chemistry and they were able to estimate the eddy diffusion parameter ($K_{zz}$) with a best fit of about log$_{10}$($K_{zz}$)=2. They recommend that treatment of non-equilibirum chemistry is necessary to capture the atmosphere of this object. 

\par JWST NIRSpec gives us access to moderate resolution spectra from 3--5 $\mu$m, where can can further investigate disequilibrium effects. Our adopted best fit temperature is 1103 K with a range of 1000--1200 K, the surface gravity is 4.5 dex with a range of 4.14--5.0, and the best fit log$_{10}$($K_{zz}$) is 5.03 with a range of 5.0--5.444. These results encompass values reported in previous studies quoted in Section \ref{hd19467b}, but our vertical mixing parameter is much higher than what \cite{greenbaum2023} found. This inconsistency could be due to the different NIRCam reduction methods\footnote{In particular, the \citet{greenbaum2023} analysis had to rely on initial photometric calibrations based on ground predictions, which have subsequently been replaced by more accurate in-flight photometric calibrations.} and the fact that the NIRSpec spectrum offers moderate resolution to model the molecular broadband features as well as the individual molecular lines. The AKARI spectra revealed the presence and strength of CO$_2$ absorption in brown dwarf atmospheres, our JWST data show individual molecular features that can constrain the abundances of the molecules in this wavelength range which is essential to capture the disequilibrium processes affecting the absorption features seen. 


Brown dwarfs are excellent objects to compare to the faint, harder to observe, high contrast exoplanets as they occupy similar color-space to planetary-mass companions \citep{faherty2013,liu2016} and have similar atmospheric processes. By observing HD 19467 B, we were able to conduct a full atmospheric analysis using a variety of public forward models and the NEWERA-PHOENIX model grid. Most brown dwarfs are expected to have an approximately Solar C/O ratio as they wouldn't have undergone core accretion formation \citep{schlaufman2018,madhu2012,oberg2011,lodders2004,madhu2011,konopacky2013}. After our analysis, we concluded that HD 19467 B indeed has a C/O ratio that is consistent with Solar as expected. 


\subsection{Evolutionary Model Comparison} \label{evol_model}

Model-independent dynamical mass measurements of substellar companions offer a rare opportunity to test and verify substellar evolutionary models that contain many assumptions about the physics of such objects including their internal structure and initial entropy. HD~19467~B is one such case in which this comparison between dynamical model-independent mass and a mass derived from its cooling curve of an evolutionary model can be made. 

In order to compare the dynamical mass of HD~19467~B, we use the \texttt{species} package \citep{stolker2020} to convert contrast values derived in \cite{Ruffio2023} across multiple JWST/NIRCam filters into evolutionary model masses using the ATMO evolutionary models for cool T-Y brown dwarfs \citep{phillips2020}. We do this for a given age, which is the adopted value in this paper of 9.4~Gyr. From doing this, we find mass values in the range from 63-75~$M_{\mathrm{Jup}}$ across the six medium filter bands as presented in \citet{Ruffio2023}. This mass range derived from the ATMO models is in agreement with our dynamical mass value derived in this paper of $71.6^{+5.3}_{-4.6}~M_{\mathrm{Jup}}$.

\subsection{Implications for Observations of Other Objects}

\par Like VHS 1256 b \citep{miles2023} and TWA 27B \citep{luhman2023}, this data showcases the ability of JWST to detect many molecular lines, and broadband features at high resolution across a wide wavelength range only accessible with JWST. We are able to see disequilibrium processes happening and how they impact the spectral features of directly imaged companions. We now can compare spectra of objects of different ages, dynamical set ups, separations, and temperatures in this wavelength regime to learn even more about the chemistry driving their atmospheres. 

But in the case of HD 19467 B, this dataset showcases JWST's ability to do high contrast spectroscopy, allowing the direct imaging community to access these molecular lines and broadband features from objects that are at closer separation (down to 0.3 arcsecs at 3 $\times$ 10$^{-5}$ and 1 arcsec at 3 $\times$ 10$^{-5}$) to their host stars \citep{Ruffio2023}. The spectrum in this paper was obtained using new high contrast processing methods that are currently shown to make spectroscopy accessible for almost all directly imaged planets, minus a few that remain close to their host star \citep{Ruffio2023}. These methods can only improve over time with iteration in methods, improvement in PSF characterization, and adapting for use on other instruments. The performance of our reduction and post-processing methods sets up the community to obtain this level of observation for the vast majority of directly imaged Jovian exoplanets. 

\section{Conclusion}\label{conclusion}


In this paper, we present the atmospheric characterization of HD 19467 B using the spectrum extracted in \cite{Ruffio2023}. HD 19467 B was chosen for this program because of its "benchmark'' status with a dynamical mass and previous studies that expect it to have stellar composition, that provides a key laboratory to test substellar evolutionary models and atmospheric models. The spectral features detected in the new spectrum were absorption from H$_2$O from about 2.9--3 $\mu$m, CH$_4$ around 3.3 $\mu$m, CO$_2$ around 4.2 $\mu$m and CO from 4.5 $\mu$m (Fig. \ref{fig:spec}). We conducted an empirical analysis using AKARI brown dwarf spectra, the only observations of substellar companions at with a similar wavelength range to JWST NIRSpec, and found the closest match to be 2MASS J0559-14, a T4.5 brown dwarf, which confirmed our spectral features and the spectral type of HD 19467 B being a late T dwarf. By confirming the spectral type and molecular features, the JWST NIRSpec data are what we expect from ground-based studies. 


JWST NIRSpec has provided the opportunity to better understand non-equilibrium chemistry and provide higher resolution data in the 3--5 $\mu$m range and put our current and new models to the test. We used three public model grids; BTSETTL08, DRIFT-PHOENIX, and Sonora Bobcat with our various best-fits reported in Table \ref{tab:atm_param}. The public grids struggled to simultaneously fit the CH$_4$ feature and the CO$_2$ feature, a problem that \cite{sorahana2013} and \cite{yamamura2010} encountered when fitting the AKARI spectral data. We used the NEWERA-PHOENIX model grid with updated chemistry and line lists to test against the new, moderate resolution NIRSpec data. NEWERA-PHOENIX fit the data best, with best fit parameters including an effective temperature of 1103 K, with an allowed range from 1000--1200 K, a surface gravity of 4.5 dex, with a range of 4.1--5.0, and a log$_{10}$($K_{zz}$) of 5.0, with a range of 5.0--5.4. These properties encompass the values found in previous studies on the object, but our results indicate deeper vertical mixing than previously thought. 
We additionally conducted a chi-squared analysis of molecular abundances and conclude HD 19467 B to have a C/O ratio that is approximately Solar, a measurement that is expected for this type of object. With the benchmark object, HD 19467 B, there have been many measurements of its dynamical mass from the ground. Here, we estimate a dynamical mass of 71.6$^{+5.3}_{-4.6}$ M$_{\mathrm{Jup}}$, which agrees with previous estimates from \cite{brandt2021} but is higher than the model derived mass from \cite{greenbaum2023}. 

HD 19467 B continues to be a great "benchmark'' object, especially for the first high contrast spectroscopy observations with JWST's NIRSpec. We confirmed findings from the ground, and discovered deeper vertical mixing in its atmosphere due to the moderate resolution spectral features. This dataset, and two companion papers, this work and \cite{Ruffio2023}, showcase what JWST can do for high contrast imaging spectroscopy and how current and past model grids perform in the new wavelength regime. These methods can only improve over time, but we show that our current methods work well and can set the stage for higher contrast systems and closer separated systems.

\section{Acknowledgments}
\label{acknowledgements}

The authors thank the anonymous referee for reviewing this work. This paper reports work carried out in the context of the JWST Telescope Scientist Team\footnote{\url{https://www.stsci.edu/~marel/jwsttelsciteam.html}} (PI: M. Mountain). Funding is provided to the team by NASA through grant 80NSSC20K0586. Based on observations with the NASA/ESA/CSA JWST, associated with program GTO-1414 (PI: Marshall Perrin), obtained at the Space Telescope Science Institute, which is operated by AURA, Inc., under NASA contract NAS 5-03127.

K.K.W.H acknowledges funding from the Giacconi Fellowship at the Space Telescope Science Institute.

This work was supported by NSF under the AAG program (grant 1614492) and HST- GO-15955.01 from the Space Telescope Science Institute, which is operated by AURA, Inc., under NASA contract NAS 5-26555.  This work also benefited from computing time allocations from the High Performance Computing (HPC) center at the University of Arizona.

This work has made use of data from the European Space Agency (ESA) mission Gaia (\url{https://www.cosmos.esa.int/gaia}), processed by the Gaia Data Processing and Analysis Consortium (DPAC, \url{https://www.cosmos.esa.int/web/ gaia/dpac/consortium}). Funding for the DPAC has been provided by national institutions, in particular, the institutions participating in the Gaia Multilateral Agreement. This research made use of the SIMBAD database and the VizieR Catalogue access tool, both operated at the CDS, Strasbourg, France. The original descriptions of the SIMBAD and VizieR services were published in \citet{wenger2000} and \citet{ochsenbein2000}.

\clearpage
\appendix

\section{Posterior Probability Distributions and MCMC Framework}\label{sec:app}

Here we discuss the details regarding our forward modeling framework for the data. The flux from the spectra are modeled using the following equation:
\begin{equation}\label{jwstmodel}
\begin{aligned}
F_M[p] ={} & \left(M\left[p^{\ast}\left(\lambda \left[1 + \frac{V_r}{c}\right]\right),\, T_\mathrm{eff}, \log g, \mathrm{[M/H]}, K_{zz} \right]\right) \\
& \ast \kappa_G(\Delta \nu_\mathrm{inst}) \times C_{F_\lambda},
\end{aligned}
\end{equation}
where $\ast$ indicates convolution. $M$ is the photospheric model of the source, dependent on effective temperature ($T_\mathrm{eff}$), surface gravity( $\log{g}$), bulk metallicity ($[M/H]$), and the single eddy diffusion coefficient ($K_{zz}$). We keep bulk metallicity fixed relative at solar ($[M/H] = 0$). $V_r$ and $c$ are the heliocentric RV and speed of light, respectively. We caution use of the RV measurement since it is not anchored to a zero-point rest velocity and the absolute wavelength calibration of NIRSpec is still being characterized. The parameter $\kappa_G(\Delta \nu_\mathrm{inst})$ is the spectrograph line spread function (LSF), modeled as a normalized Gaussian. Lastly, we include a multiplicative flux parameter, $C_{F_\lambda}$. In the case of the flux calibrated spectra, this parameter represents the dilution factor, given by the ratio of the radius and distance, i.e., $C_{F_\lambda} \approx (R \cdot d^{-1})^2$. In the case of the continuum-subtracted data, this parameter encompasses small differences in the flux between the continuum-subtracted model and data.

To model the continuum subtracted spectra, we use Equation (\ref{jwstmodel}), however, we first remove the continuum, estimating the continuum using a 1-D uniform filter 80 pixels wide over the entire spectrum and dividing it from the spectrum.

To explore the parameter space we utilize the MCMC sampler \textit{emcee} \citep{foreman-mackey2013} and the log-likelihood function,

\begin{equation}
\ln L = -0.5 \times \left[\sum \left[{\frac{\mathrm{data}[p] - D[p]}{\sigma[p]}}\right]^2 + \\ \sum \ln(2\pi({\sigma[p]})^2)\right],
\end{equation}
where \(\sigma\) is the input uncertainties, data\([p]\) is the input data, and \(D[p]\) is the forward-modeled data. We include an additional parameter on the total noise given as $\sigma = \sigma_\mathrm{inst}*C_\mathrm{Noise}$, where $\sigma_\mathrm{inst}$ represents the noise calculated from the science calibration pipeline and the multiplicative factor $C_\mathrm{Noise}$ is fit to the optimal scale so the noise encompasses the residuals between the model and the data. This allows us to model potential systematic errors not accounted for in the data or models. The uncertainty upper limit is the difference between the 84th and 50th percentile, and the lower limit for all model parameters is the difference between the 50th and 16th percentile. If the posterior distributions are Gaussian, then this simplifies to the 1-\(\sigma\) uncertainty for each parameter (e.g., \citealt{blake2010, burgasser2016}). For each modelset used we provide the grids' priors and prior distributions used for the fitting discussed in the main body of the text and shown in Table \ref{tab:param_ranges}.

\begin{figure*}
\centering
  \includegraphics[width=7in]{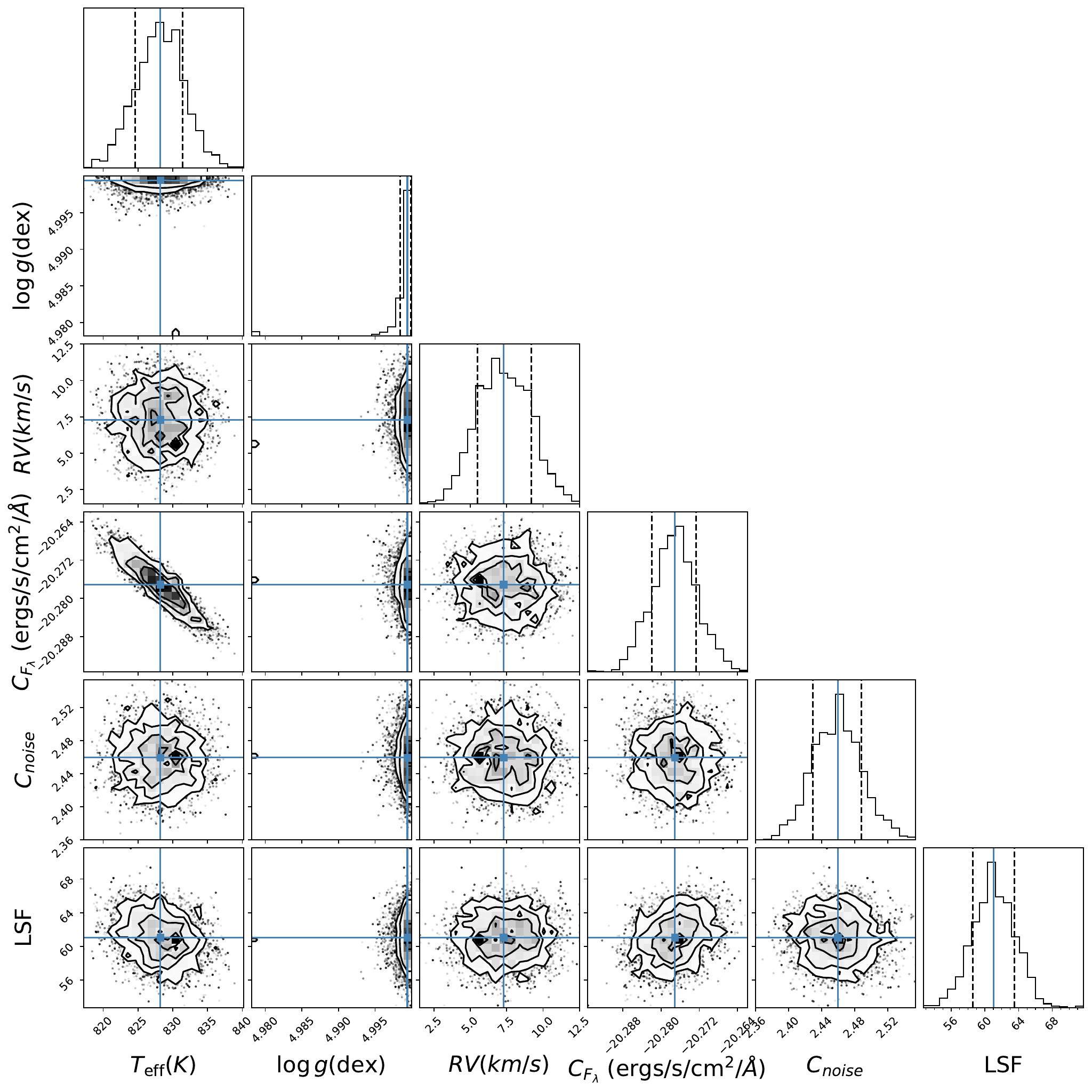}
\caption{Corner plot from our MCMC fits using the BTSETTL08 grid for our continuum NIRSpec spectra. The diagonal shows the marginalized posteriors. The covariances between all the parameters are in the corresponding 2-d histograms. The blue lines represent the 50th percentile, and the dotted lines represent the 16 and 84 percentiles. The C$_{F_\lambda}$ corresponds to the dilution factor that scales the model by \((radius)^2 (distance)^{-2}\) and C$_{\mathrm{Noise}}$ is a scaled noise parameter.}
\label{fig:btsettl08_tri}
\end{figure*}

\begin{figure*}
\centering
  \includegraphics[width=7in]{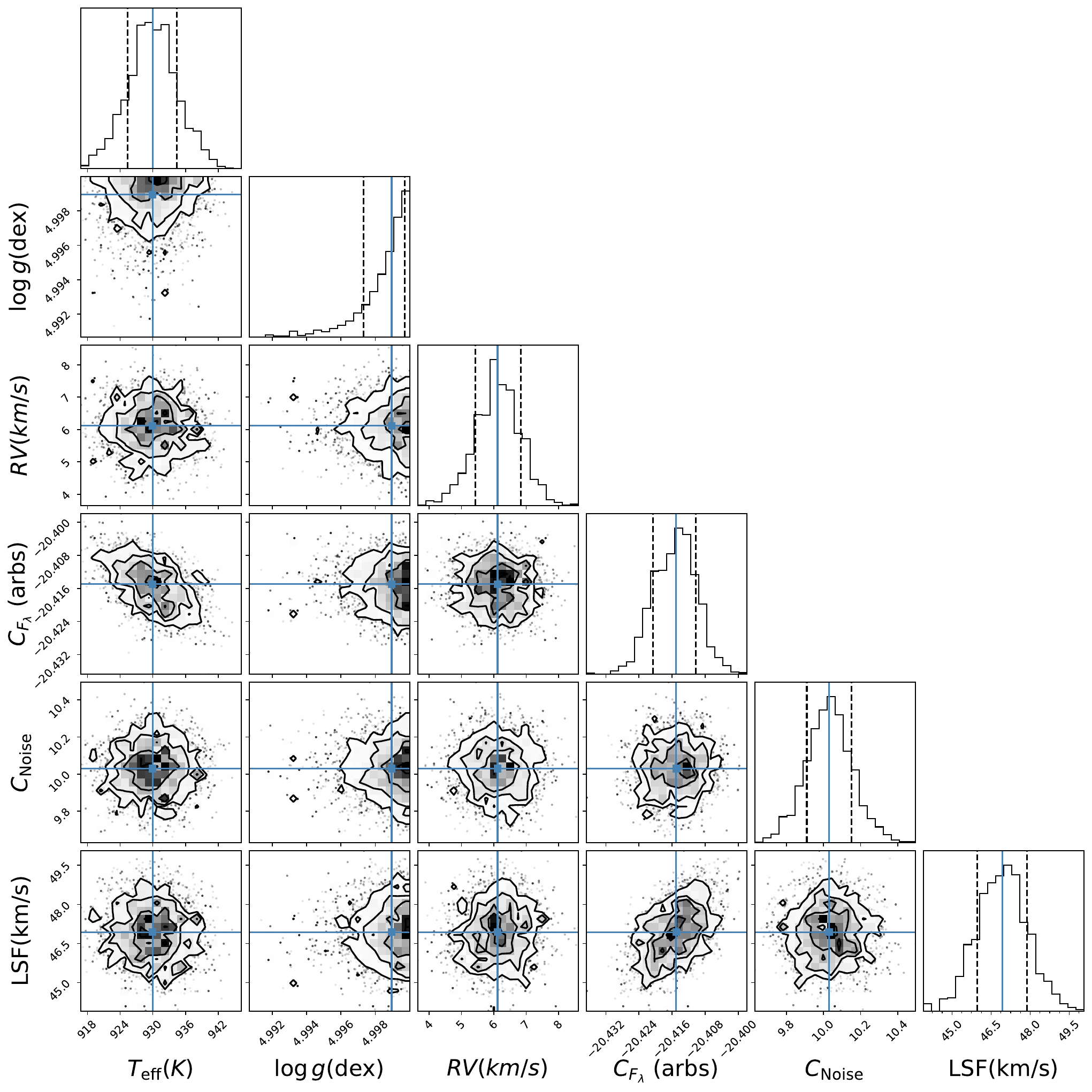}
\caption{Same as Figure~\ref{fig:btsettl08_tri} plotting our MCMC fits using the BTSETTL08 grid for our continuum-subtracted NIRSpec spectra. }
\label{fig:btsettl08_tri_flat}
\end{figure*}

\begin{figure*}
\centering
  \includegraphics[width=7in]{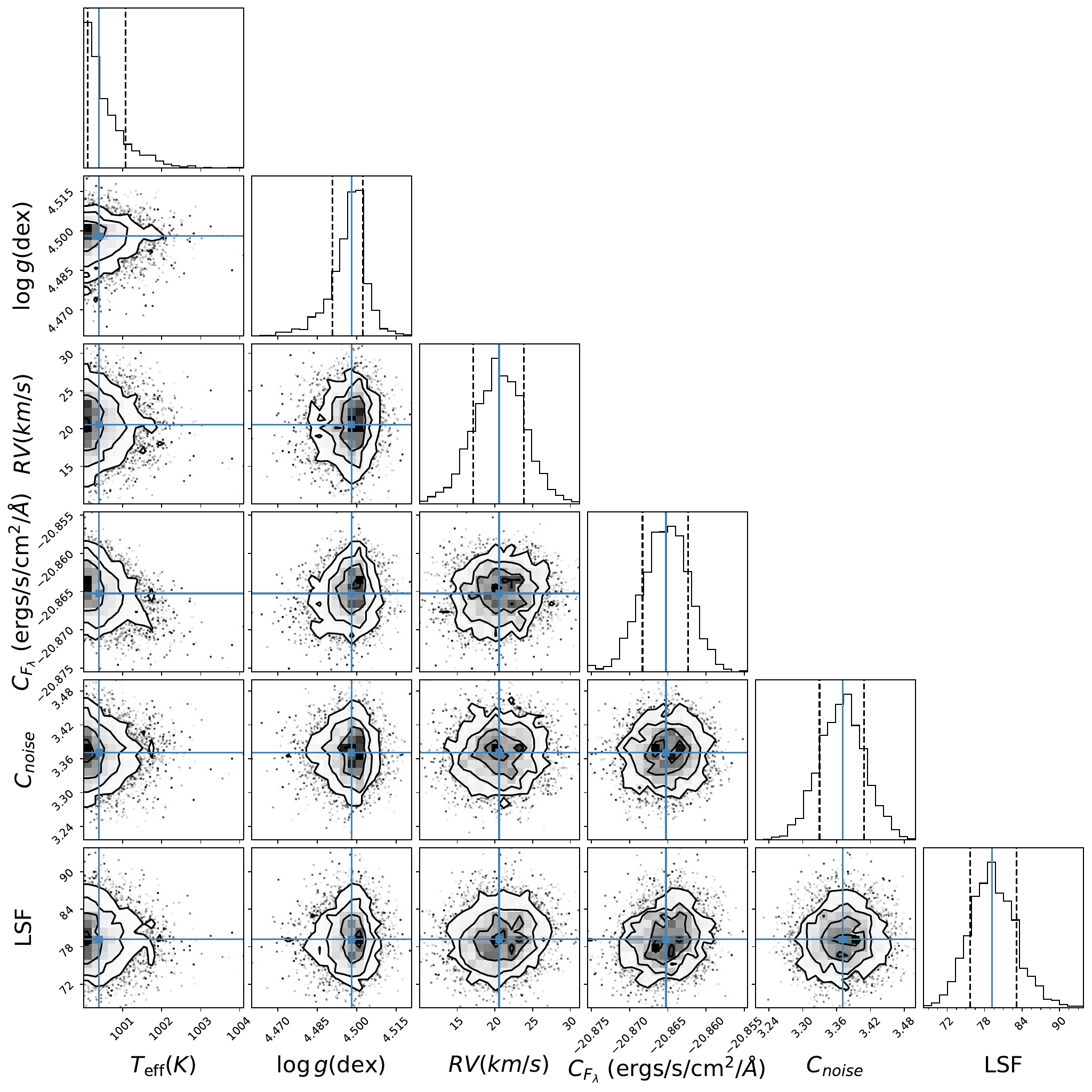}
\caption{Same as Figure~\ref{fig:btsettl08_tri} plotting our MCMC fits using the DRIFT-PHOENIX grid for our continuum NIRSpec spectra.}
\label{fig:drift_tri}
\end{figure*}

\begin{figure*}
\centering
  \includegraphics[width=7in]{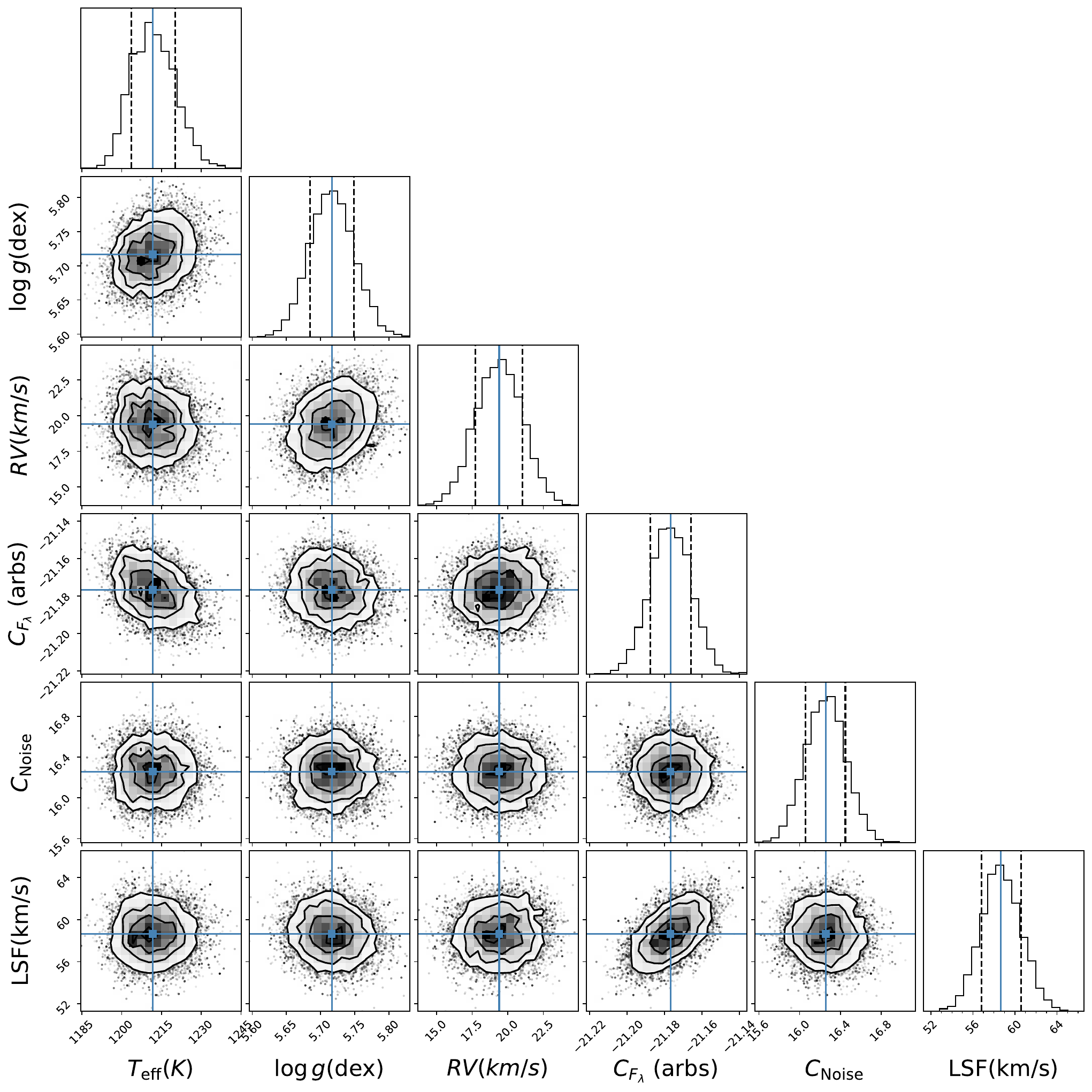}
\caption{Same as Figure~\ref{fig:btsettl08_tri} plotting our MCMC fits using the DRIFT-PHOENIX grid for our continuum-subtracted NIRSpec spectra.}
\label{fig:drift_flat_tri}
\end{figure*}

\begin{figure*}
\centering
  \includegraphics[width=7in]{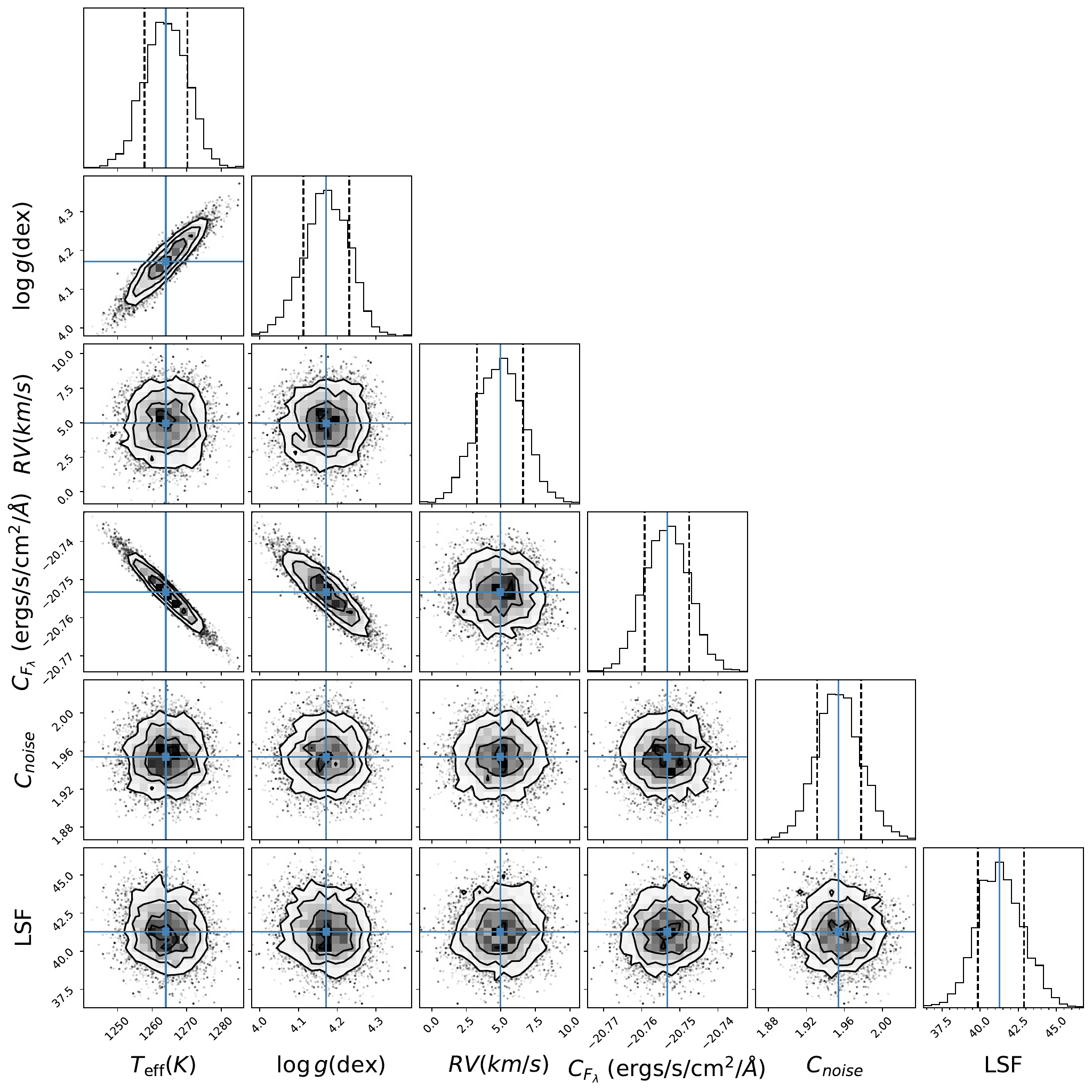}
\caption{Same as Figure~\ref{fig:btsettl08_tri} plotting our MCMC fits using the Sonora Bobcat grid for our continuum NIRSpec spectra.}
\label{fig:sonora_tri}
\end{figure*}

\begin{figure*}
\centering
  \includegraphics[width=7in]{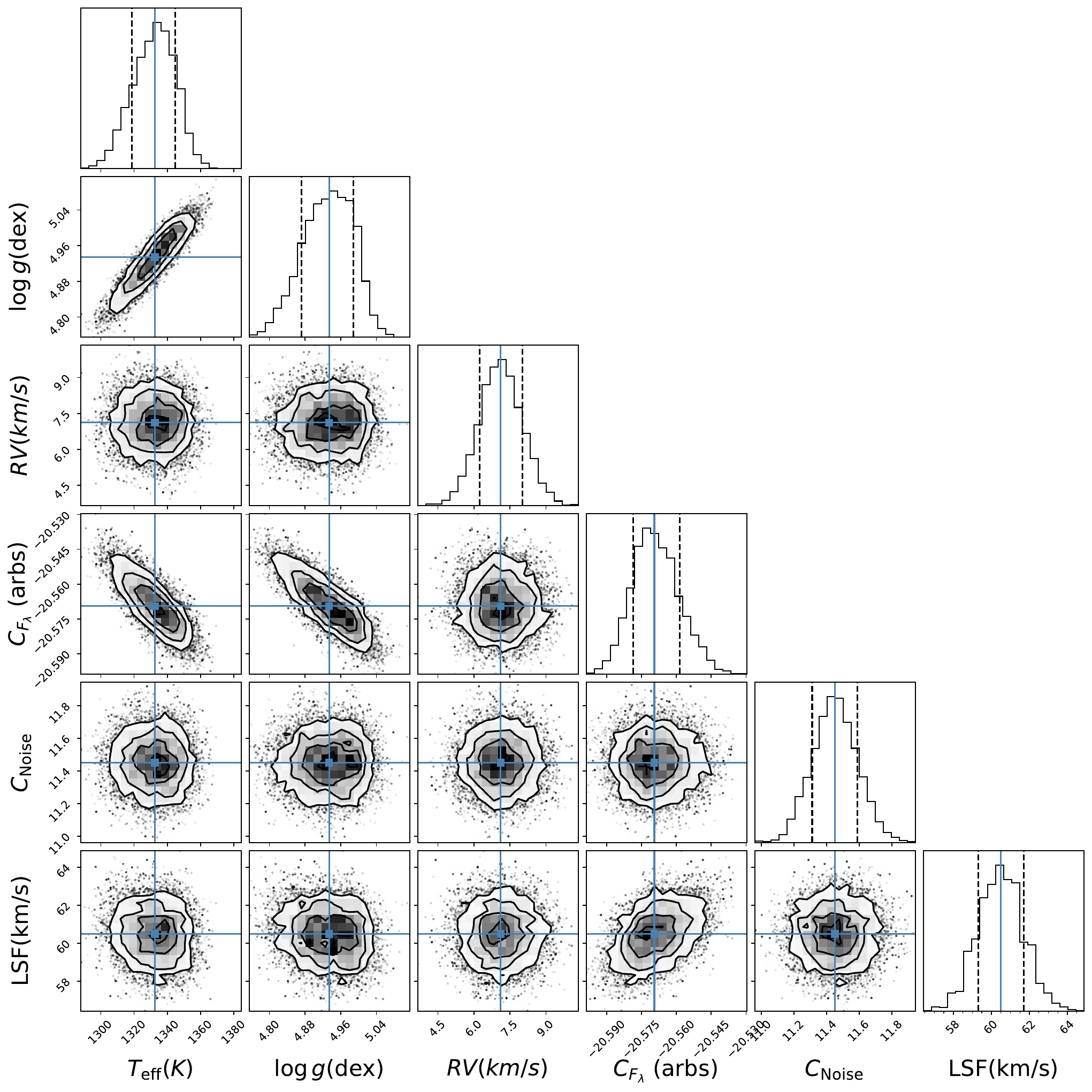}
\caption{Same as Figure~\ref{fig:btsettl08_tri} plotting our MCMC fits using the Sonora Bobcat grid for our continuum-subtracted NIRSpec spectra.}
\label{fig:triangle_sonora_flat}
\end{figure*}

\begin{figure*}
\centering
  \includegraphics[width=7in]{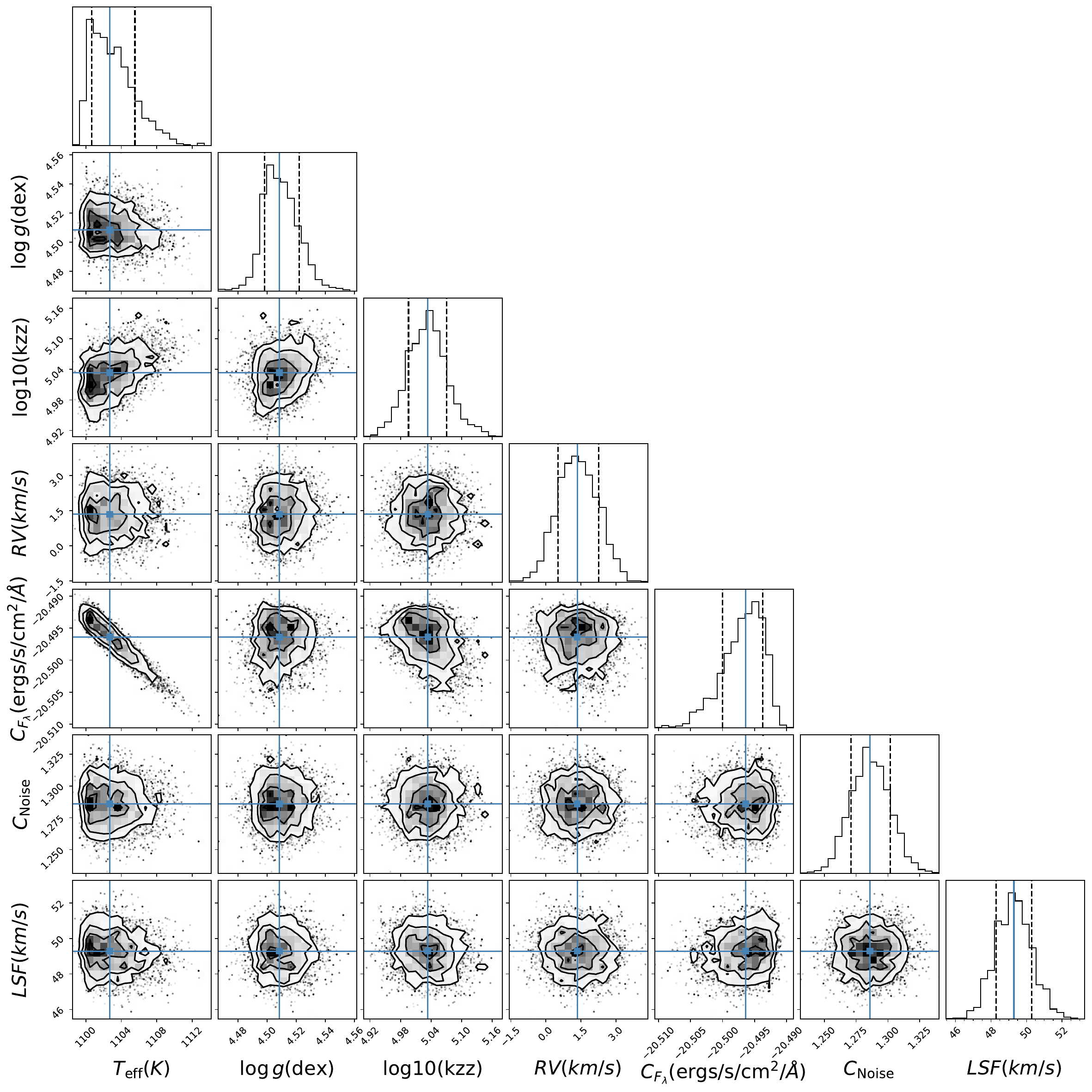}
\caption{Same as Figure~\ref{fig:btsettl08_tri} plotting our MCMC fits using the NEWERA-PHOENIX model grid for our continuum NIRSpec spectra.}
\label{fig:corner_new_era}
\end{figure*}

\begin{figure*}
\centering
  \includegraphics[width=7in]{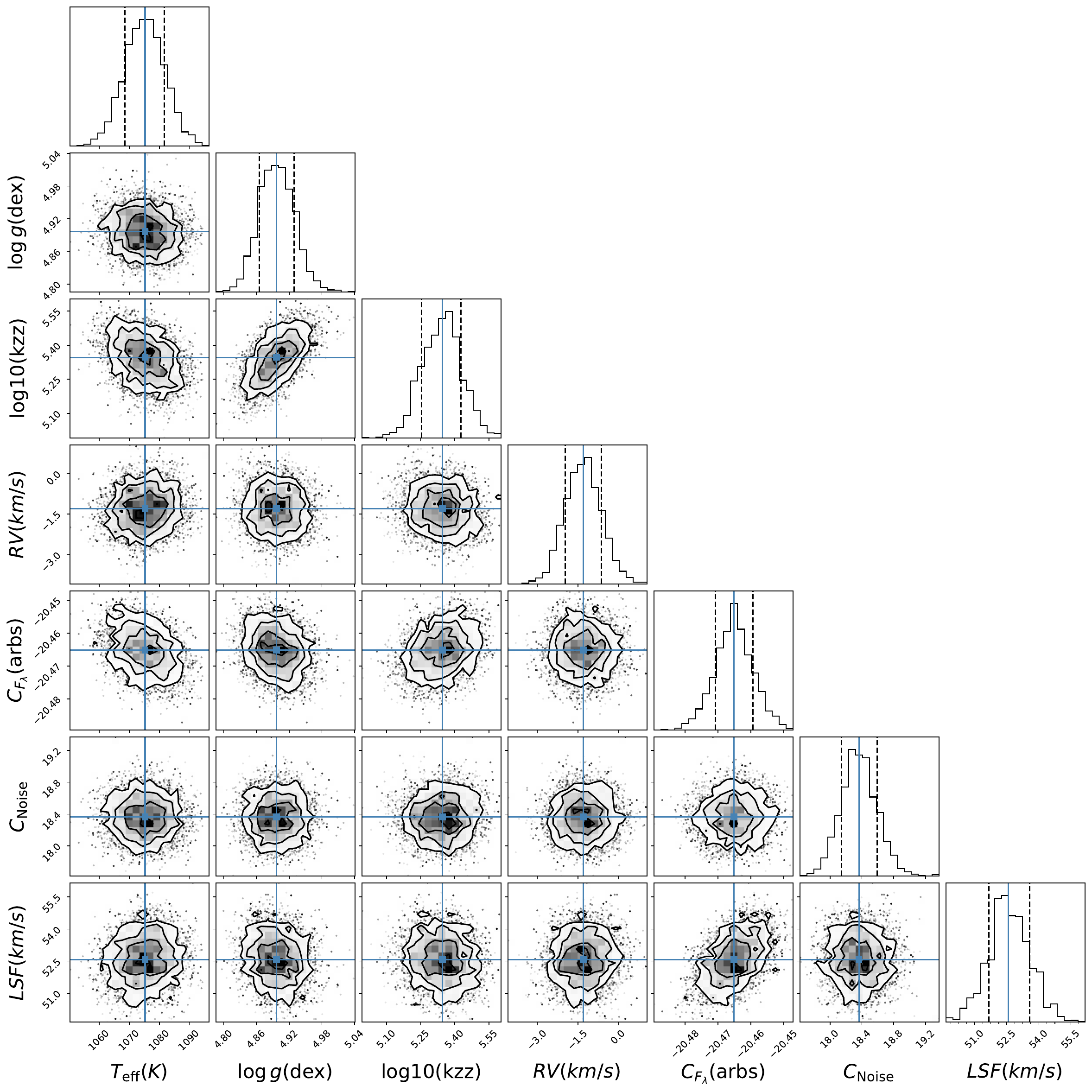}
\caption{Same as Figure~\ref{fig:btsettl08_tri} plotting our MCMC fits using the NEWERA-PHOENIX model grid for our continuum-subtracted NIRSpec spectra.}
\label{fig:corner_new_era_flat}
\end{figure*}

We provide the statistical posteriors and marginal distributions for each MCMC fit, often referred to as ``triangle" or ``corner" plots, shown in Figures~\ref{fig:btsettl08_tri}--\ref{fig:corner_new_era_flat}. These include first the posterior distributions for the atmosphere grid forward model fits in section \ref{forward_modeling}, and lastly the posterior distributions for the orbit fit in section \ref{dyn_mass}. We ensured all chains were converged and mixed using the Gelman-Rubin statistic \citep{gelman:1992:457, brooks:1998:434}. Every set of chains for each modelset had a multivariate potential scale reduction factor (MPSRF), $\hat{R}^p$, value of $<$ 1.3, indicating convergence \citep{brooks:1998:434}.

\begin{deluxetable*}{lcc}[!ht] 
\tabletypesize{\scriptsize} 
\tablewidth{\textwidth} 
\tablecaption{Forward-modeled Parameters and Priors.}
\label{tab:param_ranges}
\tablehead{ 
  \colhead{Description} & \colhead{Symbol} & \colhead{Prior (Distribution)}
} 
\startdata
\multicolumn{3}{c}{BTSETTL08 under 900 K} \\
\hline
Effective Temperature & $T_\mathrm{eff}$ & [500 -- 900] K (log-uniform) \\
Surface Gravity & $\log g$ & [3.5 -- 5.0] dex (uniform) \\
Radial Velocity & RV & [$-$1000 -- 1000] km/s (uniform) \\
Flux Multiplier\tablenotemark{a} &  $C_{F_{\lambda}}$ & [$10^{-50}$ -- $10^{-1}$]  erg/s/cm$^2$/\AA\ (uniform) \\
Noise Factor & $C_\mathrm{Noise}$ & [0.1 -- 100] (uniform) \\
LSF & $\Delta\nu_\mathrm{inst}$ & [1 -- 200] km/s (uniform) \\
\hline
\multicolumn{3}{c}{BTSETTL08 over 900 K} \\
\hline
Effective Temperature & $T_\mathrm{eff}$ & [900 -- 3500] K (log-uniform) \\
Surface Gravity & $\log g$ & [3.5 -- 5.5] dex (uniform) \\
Radial Velocity & RV & [$-$1000 -- 1000] km/s (uniform) \\
Flux Multiplier\tablenotemark{a} &  $C_{F_{\lambda}}$ & [$10^{-50}$ -- $10^{-1}$]  erg/s/cm$^2$/\AA\ (log-uniform) \\
Noise Factor & $C_\mathrm{Noise}$ & [0.1 -- 100] (uniform) \\
LSF & $\Delta\nu_\mathrm{inst}$ & [1 -- 200] km/s (uniform) \\
\hline
\multicolumn{3}{c}{Sonora Bobcat} \\
\hline
Effective Temperature & $T_\mathrm{eff}$ & [200 -- 2400] K (log-uniform) \\
Surface Gravity & $\log g$ & [3.5 -- 5.5] dex (uniform) \\
Radial Velocity & RV & [$-$1000 -- 1000] km/s (uniform) \\
Flux Multiplier\tablenotemark{a} &  $C_{F_{\lambda}}$ & [$10^{-50}$ -- $10^{-1}$]  erg/s/cm$^2$/\AA\ (log-uniform) \\
Noise Factor & $C_\mathrm{Noise}$ & [0.1 -- 100] (uniform) \\
LSF & $\Delta\nu_\mathrm{inst}$ & [1 -- 200] km/s (uniform) \\
\hline
\multicolumn{3}{c}{DRIFT-PHOENIX} \\
\hline 
Effective Temperature & $T_\mathrm{eff}$ & [1000 -- 3500] K (log-uniform) \\
Surface Gravity & $\log g$ & [3.5 -- 6.0] dex (uniform) \\
Radial Velocity & RV & [$-$1000 -- 1000] km/s (uniform) \\
Flux Multiplier\tablenotemark{a} &  $C_{F_{\lambda}}$ & [$10^{-50}$ -- $10^{-1}$]  erg/s/cm$^2$/\AA\ (log-uniform) \\
Noise Factor & $C_\mathrm{Noise}$ & [0.1 -- 100] (uniform) \\
LSF & $\Delta\nu_\mathrm{inst}$ & [1 -- 200] km/s (uniform) \\
\hline
\multicolumn{3}{c}{NEWERA-PHOENIX} \\
\hline  
Effective Temperature & $T_\mathrm{eff}$ & [800 -- 1350] K (log-uniform) \\
Surface Gravity & $\log g$ & [3.5 -- 5.5] dex (uniform) \\
Eddy Diffusion Coefficient & $K_{zz}$ & [10$^2$ -- 10$^{10}$] (log-uniform) \\
Radial Velocity & RV & [$-$1000 -- 1000] km/s (uniform) \\
Flux Multiplier\tablenotemark{a} &  $C_{F_{\lambda}}$ & [$10^{-50}$ -- $10^{-1}$]  erg/s/cm$^2$/\AA\ (log-uniform) \\
Noise Factor & $C_\mathrm{Noise}$ & [0.1 -- 100] (uniform) \\
LSF & $\Delta\nu_\mathrm{inst}$ & [1 -- 200] km/s (uniform) \\
\enddata
\tablenotetext{a}{For the continuum-subtracted data our priors are [$10^{-50}$ -- $10^{50}$] (log-uniform)}
\end{deluxetable*}

\label{app:corners}

\section{Posterior probability distributions for the orbital parameters of HD 19467 B}\label{sec:app2}

Here we show the posterior distributions of the orbital parameters for HD 19467 B that were constrained when deriving the dynamical mass of 71.6$^{+5.3}_{-4.6}$M$_{\mathrm{Jup}}$ in Section \ref{dyn_mass}. These parameters are within 1 $\sigma$ agreement with previous studies \citep{maire2020,brandt2021,greenbaum2023}.

\begin{figure*}
    \centering
    \includegraphics[width=0.98\textwidth]{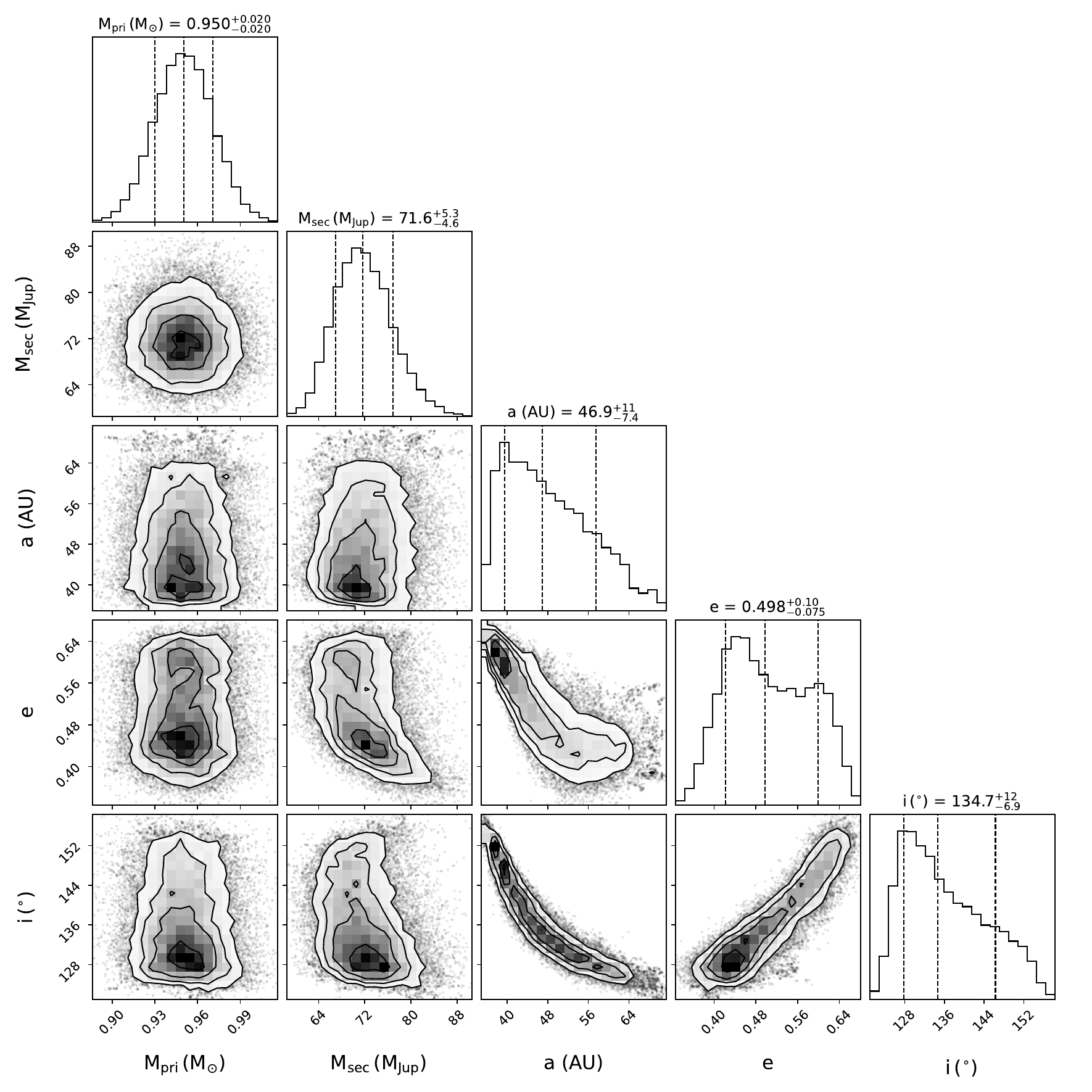}
    \caption{Marginalized 1D and 2D posterior distributions for selected orbital parameters of HD~19467~B corresponding to the fit of the RV, relative astrometry from direct imaging observations, and absolute astrometry from Hipparcos and Gaia with the use of \texttt{orvara} \citep{brandtorvara2021}. Confidence intervals at 15.85\%, 50.0\%, and 84.15\% are over-plotted on the 1D posterior distributions, with the median $\pm1~\sigma$ values given at the top of each 1D distribution. The 1, 2, and 3~$\sigma$ contour levels are over-plotted on the 2D posterior distribution.}
    \label{fig:orbit-corner-plot}
\end{figure*}

\bibliography{bibliography}

\end{document}